# Loss-induced suppression and revival of lasing

B. Peng[1†], Ş. K. Özdemir[1†*], S. Rotter[2], H. Yilmaz[1], M. Liertzer[2], F. Monifi[1], C. M. Bender[3], F. Nori[4,5], L. Yang[1*]

**Affiliations:**

[1]Department of Electrical and Systems Engineering, Washington University, St. Louis, MO 63130, USA

[2] Institute for Theoretical Physics, Vienna University of Technology, A-1040 Vienna, Austria

[3]Department of Physics, Washington University, St. Louis, MO 63130, USA

[4]Center for Emergent Matter Science, RIKEN, Wako-shi, Saitama 351-0198, Japan

[5] Physics Department, University of Michigan, Ann Arbor, Michigan 48109-1040, USA

*Correspondence to: ozdemir@ese.wustl.edu, yang@ese.wustl.edu

†These authors contributed equally to this work.

Controlling and reversing the effects of loss are major challenges in optical systems. For lasers losses need to be overcome by a sufficient amount of gain to reach the lasing threshold. We show how to turn losses into gain by steering the parameters of a system to the vicinity of an exceptional point (EP), which occurs when the eigenvalues and the corresponding eigenstates of a system coalesce. In our system of coupled microresonators, EPs are manifested as the loss-induced suppression and revival of lasing. Below a critical value, adding loss annihilates an existing Raman laser. Beyond this critical threshold, lasing recovers despite the increasing loss, in stark contrast to what would be expected from conventional laser theory. Our results exemplify the counterintuitive features of EPs and present an innovative method for reversing the effect of loss.

Dissipation is ubiquitous in nature; the states of essentially all physical systems thus have a finite decay time. A proper description of this situation requires a departure from conventional Hermitian models with real eigenvalues and orthogonal eigenstates to non-Hermitian models featuring complex eigenvalues and nonorthogonal eigenstates *(1,2,3)*. When tuning the parameters of such a dissipative system, its complex eigenvalues and the corresponding eigenstates may coalesce, giving rise to a non-Hermitian degeneracy, also called an Exceptional Point (EP) *(4)*. The presence of such an EP has a dramatic effect on the system, leading to nontrivial physics with interesting counterintuitive features such as "resonance trapping" *(5)*, a mode exchange when encircling an EP *(6)*, and a singular topology in the parameter landscape *(7)*. These characteristics can control the flow of light in optical devices with both loss and gain. In particular, waveguides having parity-time symmetry *(8)*, where loss and gain are balanced, have attracted enormous attention *(9,10)*, with effects such as loss-induced transparency *(11)*, unidirectional invisibility *(12)*, and reflectionless scattering *(13,14)* having been already observed.

Theoretical work indicates that EPs give rise to many more intriguing effects when they occur near the lasing regime; for example, enhancement of the laser linewidth *(15,16)*, fast self-pulsations *(15)*, and a pump-induced lasing death *(17)*. Realizing such anomalous phenomena, however, requires moving from waveguides to resonators, which can trap and amplify light resonantly beyond the lasing threshold. With the availability of such devices *(18,19)*, we discuss here the most counterintuitive aspect that close to an EP lasing should be inducible solely by adding loss to a resonator.

Our experimental system *(20)* consists of two directly-coupled silica whispering-gallery-mode resonators (WGMRs) $\mu R_1$ and $\mu R_2$, each coupled to a different fiber-taper WG1 and WG2 (Fig. 1A and sec. S1). The resonance frequencies of the WGMRs were tuned to be the same via thermo-optic effect, and a controllable coupling strength $\kappa$ was achieved between the WGMRs by adjusting the inter-resonator distance. To observe its behavior in the vicinity of an EP, the system was steered parametrically via $\kappa$ and an additional loss $\gamma_{\text{tip}}$ induced on $\mu R_2$ by a chromium (Cr)-coated silica-nanofiber tip (Figs. 1B and 1C), which features strong absorption in the 1550 nm band. The strength of $\gamma_{\text{tip}}$ was increased by enlarging the volume of the nanotip within the μR$_2$ mode field, resulting in a broadened resonance linewidth with no observable change in resonance frequency (Fig. 1D). A small fraction of the scattered light from the nanotip coupled back into μR$_2$ in the counter-propagating (backward) direction and led to a resonance peak whose linewidth was broadened as the loss was increased (Fig. 1E). The resonance peak in the backward direction was approximately $1/10^4$ of the input field, confirming that the linewidth-broadening and the decrease of the resonance depth in the forward direction were due to $\gamma_{tip}$ via absorption and scattering losses, but not due to back-scattering into the resonator.

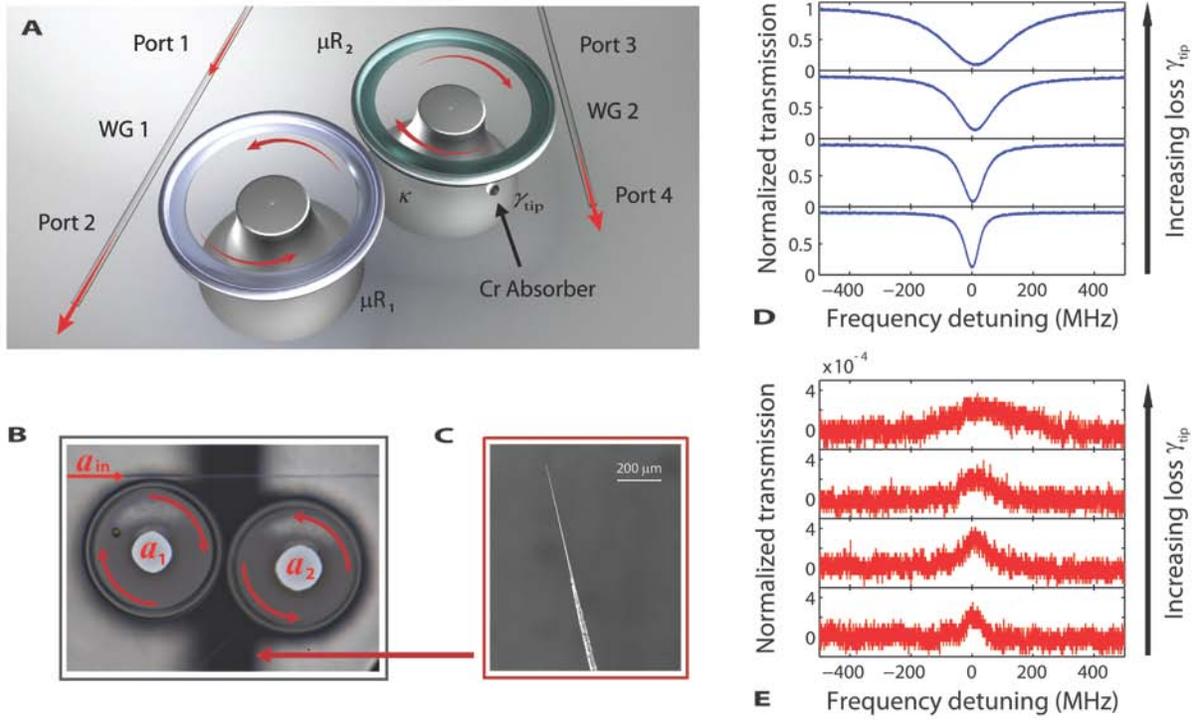

*Fig.1. Coupled WGM microresonators and the effect of loss. (A) An illustration of the coupled resonators μR$_1$ and μR$_2$ with the fiber-taper couplers WG1 and WG2. (B) Optical microscope image of the resonators with WG1 and the Cr nanotip. $a_{in}$: input field at WG1. $a_1$ & $a_2$: rotating intracavity fields of μR$_1$ & μR$_2$. κ : inter-resonator coupling strength. $\gamma_{tip}$ : additional loss induced by the nanotip. (C) Scanning electron microscope (SEM) image of the Cr nanotip. (D and E) Transmission spectra in the forward (D) and backward (E) direction.*

In the first set of experiments WG2 was moved away from μR$_2$ to eliminate the coupling between them. We investigated the evolution of the eigenfrequencies and the transmission spectra $T_{1\to 2}$ from input port 1 to output port 2 by continuously increasing $\gamma_{tip}$ while keeping $\kappa$ fixed. In this configuration, losses experienced by μR$_1$ and μR$_2$ were $\gamma'_1 = \gamma_1 + \gamma_{c1}$ and $\gamma'_2 = \gamma_2 + \gamma_{tip}$, where $\gamma_{c1}$ is the WG1-μR$_1$ coupling loss, and $\gamma_1$ and $\gamma_2$ include material absorption, scattering, and radiation losses of μR$_1$ and μR$_2$. The coupling between the WGMRs led to the formation of two supermodes with complex eigenfrequencies $\omega_{\mp} = \omega_0 - i\chi \mp \beta$ whose real and imaginary parts are respectively given by $\omega'_{\mp}$ and $\omega''_{\mp}$ *(20)*. Here $\omega_0$ is the resonance frequency of the solitary WGMRs, $\chi = (\gamma'_1 + \gamma'_2)/4$ and $\Gamma = (\gamma'_1 - \gamma'_2)/4$ respectively quantify the total loss and the loss contrast of the WGMRs, and $\beta = \sqrt{\kappa^2 - \Gamma^2}$ reflects the transition between the strong and the weak inter-mode coupling regimes due to an interplay of the inter-resonator coupling strength $\kappa$ and the loss contrast $\Gamma$ (sec. *S1*). In the strong-coupling regime, quantified by $\kappa > |\Gamma|$ and real $\beta$, the supermodes had different resonance frequencies (mode splitting of $2\beta$) but the same linewidths quantified by $\chi$. This was reflected as two spectrally-separated resonance modes in $T_{1\to 2}$ [Fig. 2A(i)] and in the corresponding eigenfrequencies [Fig. 2B(i)]. Since our system satisfied $\gamma_1 + \gamma_{c1} > \gamma_2$, introducing $\gamma_{tip}$ to μR$_2$ increased the amount of splitting until $\gamma_1 + \gamma_{c1} = \gamma_2 + \gamma_{tip}$ (that is, $\gamma'_1 = \gamma'_2$) was satisfied [Fig. 2A(ii) and 2B(ii)]. Increasing $\gamma_{tip}$ beyond this point gradually led to an overlap of the supermode resonances [Fig. 2A(iii)], such as to necessitate a fit to a theoretical model to extract the complex resonance parameters (sec. *S1&S2*) *(20)*. At $\gamma_{tip} = \gamma_{tip}^{EP}$ where $\kappa = |\Gamma|$, the supermodes coalesced at the EP. With a further increase of $\gamma_{tip}$ the system entered the weak-coupling regime, quantified by $\kappa < |\Gamma|$ and imaginary $\beta$, leading to two supermodes with the same resonance frequency but different linewidths [Fig. 2A(iv) and 2B(iv)]. The resulting resonance trajectories in the complex plane clearly displayed a reversal of eigenvalue evolution (Fig. 2B): The real parts of the eigenfrequencies of the system approached each other while their imaginary parts remained equal until the EP. After passing the EP, their imaginary parts were repelled, resulting in an increasing imaginary part for one of the eigenfrequencies and a decreasing imaginary part for the other. As a result, one of the modes became less lossy while the other became more lossy (sec. S4).

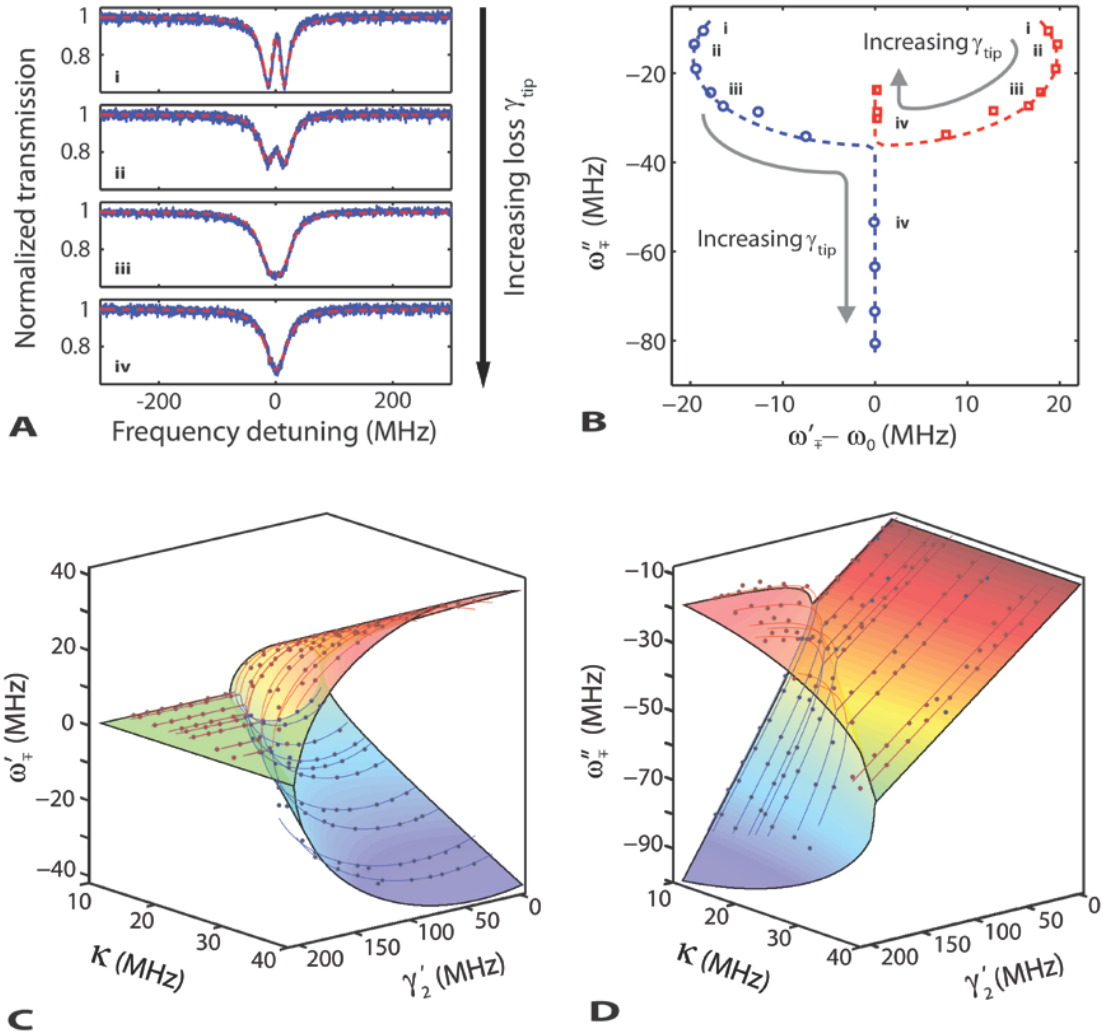

*Fig.2. Evolution of the transmission spectra and the eigenfrequencies as a function of loss $\gamma_{tip}$ and inter-resonator coupling strength $\kappa$. (A) Transmission spectra $T_{1\to 2}$ showing the effect of loss on the supermodes. Blue and red curves denote the experimental data and the best fit using a theoretical model (sec. S1&S2), respectively. (B) Evolution of the eigenfrequencies of the supermodes in the complex plane as $\gamma_{tip}$ was increased. Open circles and squares are the eigenfrequencies estimated from the measured $T_{1\to 2}$. Dashed red and blue lines denote the best theoretical fit to the experimental data. (C and D) Eigenfrequency surfaces in the ($\kappa, \gamma'_2$) parameter space (sec. S4).*

By repeating the experiments for different $\kappa$ and $\gamma_{tip}$ we obtained the eigenfrequency surfaces $\omega_{\mp}(\kappa, \gamma'_2)$ whose real and imaginary parts are shown in Figs. 2C and 2D. The resulting surfaces exhibit a complex square-root-function topology with the special feature that a coalescence of the eigenfrequencies can be realized by varying either $\kappa$ or $\gamma_{tip}$ alone, leading to a continuous thread of EPs along what may be called an exceptional line. As expected, the slope of this line is such that stronger $\kappa$ requires higher $\gamma_{tip}$ to reach the EP (sec. S4).

Our second set of experiments was designed to elucidate the effect of the EP on the intracavity field intensities. For this we used both WG1 and WG2, introducing an additional coupling loss $\gamma_{c2}$ to μR$_2$ (that is, $\gamma'_2 = \gamma_2 + \gamma_{tip} + \gamma_{c2}$). We tested two different cases by choosing different mode-pairs in the resonators *(20)*. In Case 1, the mode in μR$_1$ had higher loss than the mode in μR$_2$ ($\gamma_1 + \gamma_{c1} > \gamma_2 + \gamma_{c2}$); in Case 2, the mode in μR$_2$ had higher loss ($\gamma_1 + \gamma_{c1} < \gamma_2 + \gamma_{c2}$). The system was adjusted so that two spectrally-separated supermodes were observed in the transmission spectra $T_{1\to2}$ and $T_{1\to4}$ as resonance dips and peaks (sec. S3). No resonance dip or peak was observed at port 3. Using experimentally-obtained $T_{1\to2}$ and $T_{1\to4}$, we estimated the intracavity fields $I_1$, $I_2$ and the total intensity $I_T = I_1 + I_2$ as a function of $\gamma_{tip}$ (Fig. 3A-C and sec. S1,S2,S5&S6). As $\gamma_{tip}$ was increased, $I_T$ first decreased and then started to increase despite increasing loss. This loss-induced recovery of the intensity is in contrast to the expectation that the intensity would decrease with increasing loss and is a direct manifestation of the EP.

The effect of increasing $\gamma_{tip}$ on $I_1$ and $I_2$ at $\omega_{\mp}$ is depicted in Figs. 3A&B. When $\gamma_{tip} = 0$ and the system was set in the strong-coupling regime, the light input at μR$_1$ was freely exchanged between the resonators, establishing evenly distributed supermodes. As a result, the intracavity field intensities were almost equal. As $\gamma_{tip}$ was increased, $I_1$ and $I_2$ decreased continuously at different rates until $I_1$ reached a minimum at $\gamma_{tip} = \gamma_{tip}^{min}$. The rate of decrease was higher for $I_2$ due to increasingly higher loss of μR$_2$. Beyond $\gamma_{tip}^{min}$, until the EP was reached at $\gamma_{tip} = \gamma_{tip}^{EP}$, the system remained in the strong-coupling regime, but the supermode distributions were strongly affected by $\gamma_{tip}$, leading to an increase of $I_1$ and hence of $I_T$ while no appreciable change was observed for $I_2$ *(20)*. Increasing $\gamma_{tip}$ further pushed the system beyond the EP, thereby completing the transition from the strong-coupling to the weak-coupling regime during which $I_1$ increased and kept increasing whereas $I_2$ continued decreasing. This behavior is a manifestation of the progressive localization of one of the supermodes in the less lossy μR$_1$ and of the other supermode in the more lossy μR$_2$. We conclude that the non-monotonic evolution of $I_T$ for increasing values of $\gamma_{tip}$ is the result of a transition from a symmetric to an asymmetric distribution of the supermodes in the two resonators (sec. S5&S7).

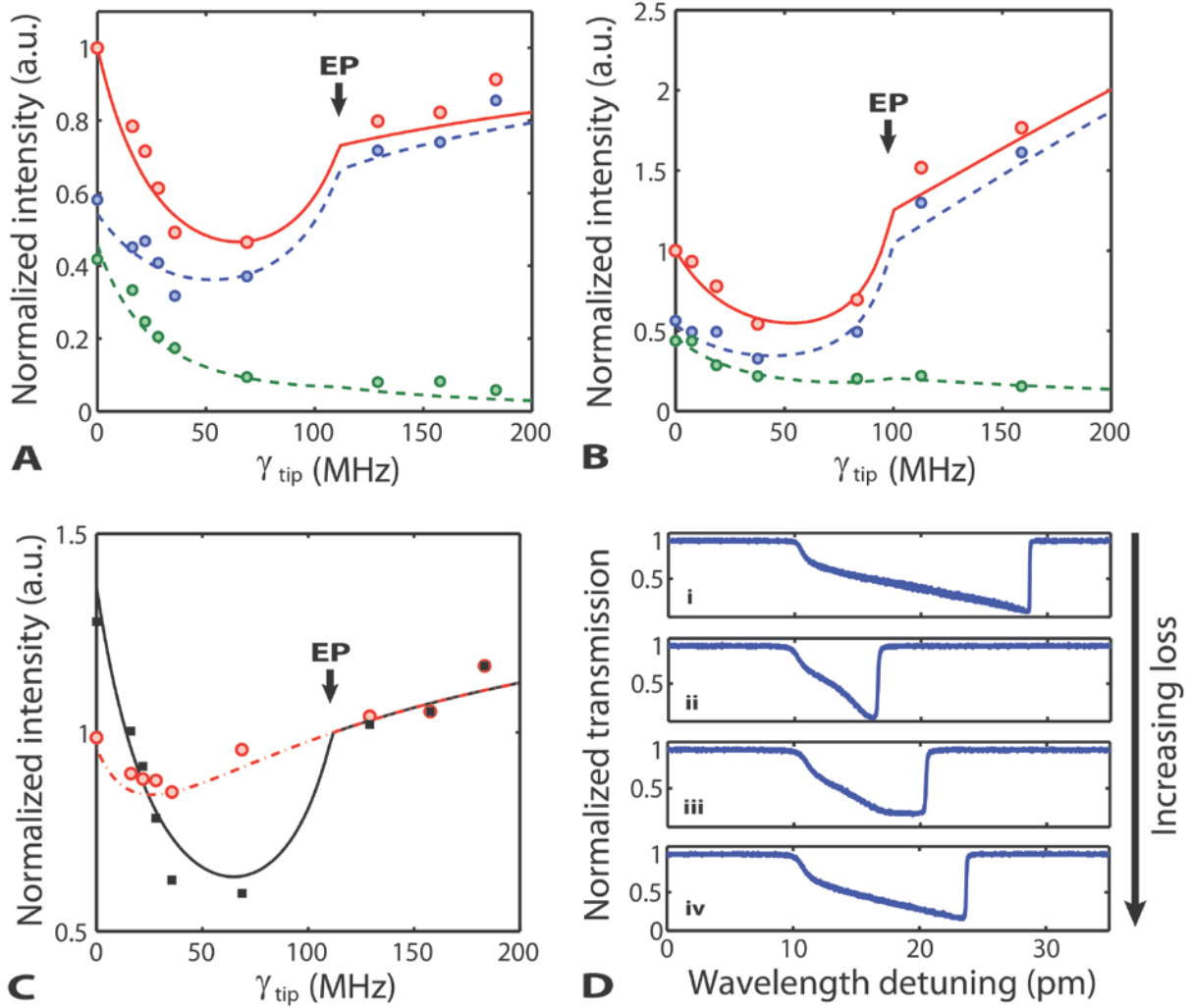

*Fig.3. Loss-induced enhancement of intracavity field intensities and thermal nonlinearity in the vicinity of an exceptional point. (A&B) Intracavity field intensities of the resonators at $\omega_\mp$ (blue: $I_1$ of $\mu R_1$, green: $I_2$ of $\mu R_2$, and red: total $I_T$) for (A) Case 1, and (B) Case 2. Normalization was done with respect to the total intensity at $\gamma_{tip}=0$. (C) Total intracavity field intensities $I_T$ at eigenfrequencies $\omega_\mp$ (black) and $\omega_0$ (red) for Case 1 (sec. S5 for Case 2). Normalization is done with respect to the intensity at the EP. (D) Effect of loss on nonlinear thermal response of coupled resonators (sec. S8): (i) solitary resonator, (ii) coupled resonators with $\gamma_{tip}=0$, and (iii) & (iv) coupled resonators with increasing $\gamma_{tip}$(20). Circles in (A,B&C) and squares in (C) were calculated from experimentally obtained transmissions $T_{1\to2}$ and $T_{1\to4}$ whereas solid and dashed curves are from the theoretical model (sec.S1-S3).*

The data shown in Figs. 3A&B also demonstrate that the initial loss contrast of the resonators affects both the amount of $\gamma_{tip}$ required to bring the system to the EP and the intensity values themselves *(20)*: Increasing $\gamma_{tip}$ in Case 2 increased $I_T$ to a higher value than that at $\gamma_{tip}=0$; in Case 1, on the other hand, $I_T$ stayed below its initial value at $\gamma_{tip}=0$. Finally, Fig. 3C shows that the intracavity field intensities at $\omega_\pm$ and $\omega_0$ coincide when $\gamma_{tip} \geq \gamma_{tip}^{EP}$, that is, after the EP transition to the weak coupling regime *(20)*. This is a direct consequence of the coalescence of eigenfrequencies $\omega_\mp$ at $\omega_0$ (sec.S5).

Whispering-gallery-mode microresonators *(21)* combine high quality factor $Q$ (long photon storage time; narrow linewidth) and high finesse $F$ (strong resonant power build-up) with microscale mode volume $V$ (tight spatial confinement; enhanced resonant field intensity) and are thus ideal for studying quantum electrodynamics *(22)*, optomechanics *(23)*, lasing (24), and sensing *(25,26,27)*. The ability of WGMRs to provide high intracavity field intensity and long interaction time reduces thresholds for nonlinear processes. Therefore, loss-induced reduction and the recovery of intracavity field intensities should impact directly on the thermal nonlinearity *(28)* and the Raman lasing *(24, 29)* in WGMRs.

Thermal nonlinearity in WGMRs is due to the temperature-dependent resonance-frequency shifts caused by material absorption of the intracavity field and the resultant heating *(20,28)*. In silica

WGMRs this is manifested as thermal broadening (narrowing) of the resonance line when the wavelength of a probe laser is scanned from shorter (longer) to longer (shorter) wavelengths. In our system thermal nonlinearity was observed in $T_{1\to 2}$ as a shark-fin feature (Fig. 3D). With an input power of 600 µW, thermal broadening kicked in and made it impossible to resolve the individual supermodes [Fig. 3D (i)&(ii)]. When $\gamma_{tip}$ was introduced and gradually increased, thermal nonlinearity and the associated linewidth broadening gradually recovered [Fig. 3D (iii)&(iv)]. This aligns well with the evolution of the total intracavity field as a function of loss (sec. S8).

Finally, we tested the effect of the loss-induced recovery of the intracavity field intensity on Raman lasing in silica microtoroids *(29,30)*. The threshold for Raman lasing scales as $P_{\text{Raman-threshold}} \propto V/g_R Q^2$, implying the significance of the pump intracavity field intensity in the process. With a pump in the 1550 nm band, Raman lasing in silica WGMR takes place in the 1650 nm band. Figure 4 depicts the spectrum and the efficiency of Raman lasing in our system. The lasing threshold for µR$_1$ was about 150 µW (Fig. 4B blue curve). Keeping the pump power fixed, we introduced µR$_2$, which had a much larger loss than µR$_1$. This effectively increased the total loss of the system and annihilated the laser (Fig. 4A, gray curve). Introducing $\gamma_{tip}$ to µR$_2$ helped to recover the Raman laser, whose intensity increased with increasing $\gamma_{tip}$ (Fig. 4A). We also checked the lasing threshold of each of the cases depicted in Fig. 4A and observed that as $\gamma_{tip}$ was increased, the $P_{\text{Raman-threshold}}$ increased at first but then decreased (Fig. 4B).

These observations are in stark contrast with what one would expect in conventional systems, where the higher the loss, the higher the lasing threshold. Surprisingly, in the vicinity of an EP, less loss is detrimental and annihilates the process of interest; more loss is good because it helps to recover the process. This counterintuitive effect happens because the supermodes of the coupled system readjust themselves as loss is gradually increased. When the loss exceeds a critical value, one supermode is mostly located in the subsystem with less loss and thus the total field can build up more strongly *(20)*. As our results demonstrate, this behavior also affects nonlinear processes, such as thermal broadening and Raman lasing, that rely on intracavity field intensity.

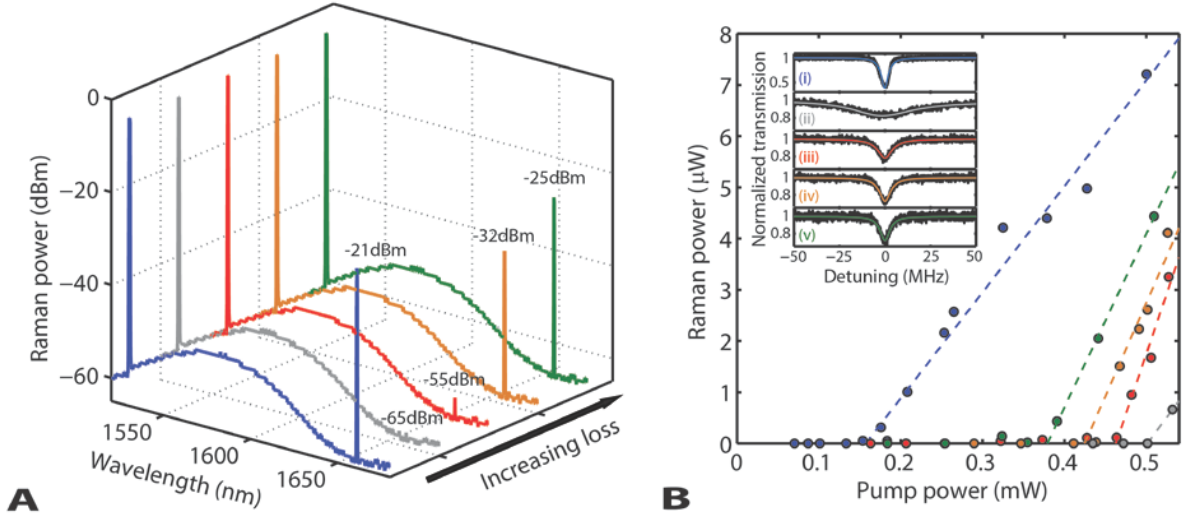

*Fig.4. Loss-induced suppression and revival of Raman lasing in silica microcavities. (A) Raman lasing spectra of coupled silica microtoroid resonators as a function of increasing loss. (B) Effect of loss on the threshold of Raman laser and its output power. The inset shows the normalized transmission spectra $T_{1\to 2}$ in the pump band obtained at very weak powers for different amounts of additional loss. Loss increases from top to bottom. The curves with the same color code in (A), (B) and the inset of (B) are obtained at the same value of additionally introduced loss.*

Our system provides a comprehensive platform for further studies of EPs and opens up new avenues of research on non-Hermitian systems and their behavior. Our findings may also lead to new schemes and techniques for controlling and reversing the effects of loss in other physical systems, such as in photonic crystal cavities, plasmonic structures, and metamaterials.

**References and Notes:**

1. C. M. Bender, Making sense of non-Hermitian Hamiltonians, *Rep. Prog. Phys.* **70**, 947 (2007).
2. I. Rotter, A non-Hermitian Hamilton operator and the physics of open quantum systems, *J. Phys. Math. Theor.* **42**, 153001 (2009).

3. N. Moiseyev, Non-Hermitian Quantum Mechanics, Cambridge University Press, (2011).

4. W. D. Heiss, Exceptional points of non-Hermitian operators, *J. Phys. A* **37**, 2455 (2004).

5. E. Persson, I. Rotter, H. J. Stöckmann, M. Barth, Observation of resonance trapping in an open microwave cavity, *Phys. Rev. Lett.* **85**, 2478–2481 (2000).

6. C. Dembowski et al., Experimental observation of the topological structure of exceptional points, *Phys. Rev. Lett.* **86**, 787–790 (2001).

7. S. B. Lee et al., Observation of an exceptional point in a chaotic optical microcavity, *Phys. Rev. Lett.* **103**, 134101 (2009).

8. C. M. Bender, S. Boettcher, Real spectra in non-Hermitian Hamiltonians having PT symmetry, *Phys. Rev. Lett.* **80**, 5243–5246 (1998).

9. R. El-Ganainy, K. G. Makris, D. N. Christodoulides, Z. H. Musslimani, Theory of coupled optical PT-symmetric structures, *Opt. Lett.* **32**, 2632 (2007).

10. C. E. Rüter et al., Observation of parity–time symmetry in optics, *Nat. Phys.* **6**, 192–195 (2010).

11. A. Guo et al., Observation of PT-symmetry breaking in complex optical potentials, *Phys. Rev. Lett.* **103**, 093902 (2009).

12. A. Regensburger et al., Parity-time synthetic photonic lattices, Nature 488, 167–171 (2012).

13. L. Feng et al., Experimental demonstration of a unidirectional reflectionless parity-time metamaterial at optical frequencies, *Nat. Mater.* **12**, 108–113 (2013).

14. L. Feng et al., Demonstration of a large-scale optical exceptional point structure, *Opt. Express* **22**, 1760–1767 (2014).

15. H. Wenzel, U. Bandelow, H.-J. Wunsche, J. Rehberg, Mechanisms of fast self pulsations in two-section DFB lasers, *IEEE J. Quantum Electron.* **32**, 69 –78 (1996).

16. M. V. Berry, Mode degeneracies and the Petermann excess-noise factor for unstable lasers, *J. Mod. Opt.* **50**, 63–81 (2003).

17. M. Liertzer et al., Pump-induced exceptional points in lasers, *Phys. Rev. Lett.* **108**, 173901 (2012).

18. B. Peng et al., Parity-time-symmetric whispering-gallery microcavities, *Nat. Phys.* **10**, 394–398 (2014).

19. M. Brandstetter et al., Reversing the pump-dependence of a laser at an exceptional point, *Nat. Commun.* **5**, 4034 (2014).

20. Supplementary materials are available on Science Online.

21. K. J. Vahala, Optical microcavities. *Nature* **106**, 839-846 (2003).

22. D. O'Shea, C. Junge, J. Volz, and A. Rauschenbeutel, Fiber-optical switch controlled by a single atom. *Phys. Rev. Lett.* **111**, 193601 (2013).

23. T. J. Kippenberg, K. J. Vahala, Cavity optomechanics: Back-action at the mesoscale. *Science* **321**, 1172-1176 (2008).

24. L. He, S. K. Ozdemir, L. Yang, Whispering gallery microcavity lasers. *Laser &Photon. Rev.* **7**, 60 (2013).

25. X. Fan, I. M. White, S. I. Shopova, H. Zhu, J. D. Suter, Y. Sun, Sensitive optical biosensors for unlabeled targets: A review. *Anal. Chim. Acta*. **620**, 8– 26 (2008).

26. F. Vollmer, S. Arnold, Whispering-gallery-mode biosensing: label-free detection down to single molecules. *Nat. Meth.* **5**, 591–596 (2008).

27. J. Zhu *et al.,* Single nanoparticle detection and sizing by mode-splitting in an ultra-high-*Q* microtoroid resonator. *Nat. Photon.* **4**, 46-49 (2010).

28. T. Carmon, L. Yang, K. J. Vahala, Dynamical thermal behavior and thermal self-stability of microcavities. *Opt. Express* **12**, 4742–4750 (2004).

29. S. Spillane, T. J. Kippenberg, K. J. Vahala, Ultralow-threshold Raman laser using a spherical dielectric microcavity. *Nature* **415**, 621 –623 (2002).

30. S. K. Ozdemir *et al.,* Highly sensitive detection of nanoparticles with a self-referenced and self-heterodyned whispering-gallery Raman microlaser. *Proc. Natl. Acad. Sci. USA*, 111, E3836-E3844 (2014).

**Acknowledgments:** SKO and LY conceived the idea and designed the experiments; BP and SKO performed the experiments with help from HY and FM. Theoretical background and simulations were provided by BP, SKO, SR, ML, CMB, and FN. All authors discussed the results, and SKO, SR and LY wrote the manuscript with inputs from all authors. This work was supported by ARO grant No. W911NF-12-1-0026. CMB was supported by DOE grant No. DE-FG02-91ER40628. FN is partially supported by the RIKEN iTHES Project, MURI Center for Dynamic Magneto-Optics, Grant-in-Aid for Scientific Research (S). SR is supported by the Vienna Science and Technology Fund project No. MA09-030 and by the Austrian Science Fund project No. SFB-IR-ON F25-P14, SFB-NextLite F49-P10.

# Supplementary Materials

## S1. Theoretical model for coupled whispering-gallery-mode (WGM) resonators.

A schematic illustration of the complete experimental setup and the configurations used in our experiments and theoretical models are given in **Fig.S1**. Here, we present a theoretical model and derive the necessary relations and expressions which will help to understand the experimental results presented in the main text. In the following we consider the general case (**Fig. S1A & S1B**) where two WGM resonators labeled as μR$_1$ and μR$_2$ are coupled to each other directly, and each of the resonators are coupled to a different fiber-taper coupler, WG1 and WG2, to couple light into and out of the WGMs. Note that in this general case (**Fig. S1A & S1B**) the total loss experienced by the resonators includes the fiber-resonator coupling losses $\gamma_{c1}$ and $\gamma_{c2}$, respectively for μR$_1$ and μR$_2$. The model and the related relations for the case where only μR$_1$ is coupled to the fiber-taper WG1 (there is no coupling between μR$_2$ and WG2) (**Fig. S1C**) can be found from the expressions below by setting $\gamma_{c2} = 0$.

Defining the intracavity mode fields (rotating waves) of the resonators as $a_{k=1,2}$ for the first and second resonators with resonance frequencies $\omega_{k=1,2}$, the coupling strength between the resonators as $\kappa$, and the input field at port 1 of the WG1 as $a_{in}$, we can write the following rate equations for the coupled-resonators system,

$$\frac{da_1}{dt} = -i\omega_1 a_1 - \frac{\gamma_1 + \gamma_{c1}}{2} a_1 - i\kappa a_2 - \sqrt{\gamma_{c1}} a_{in}$$

$$\frac{da_2}{dt} = -i\omega_2 a_2 - \frac{\gamma_2 + \gamma_{c2} + \gamma_{tip}}{2} a_2 - i\kappa a_1$$

(S.1)

together with the input-output relations $a_{out2} = a_{in} + \sqrt{\gamma_{c1}} a_1$ and $a_{out4} = \sqrt{\gamma_{c2}} a_2$ *(31)*. Here $\gamma_1$ and $\gamma_2$ denote the loss of the resonators (including material absorption, scattering, radiation and bending losses but not the coupling losses), $\gamma_{c1}$ and $\gamma_{c2}$ correspond to the losses due to the coupling of the resonators with the fiber tapers (i.e., the μR$_1$-WG1 and μR$_2$-WG2 systems), and $\gamma_{tip}$ denotes the additional loss induced by a chromium (Cr) coated nanofiber tip to the second resonator μR$_2$. In Eq. (S.1), the loss terms containing the intracavity field amplitudes $a_1$ and $a_2$ and the incoupling term containing the incoming amplitude $a_{in}$ have the same sign *(31)*. Note that the sign of the prefactor multiplying $a_{in}$ does not change the outcome of the calculations as we deal only with transmitted and reflected intensities for which the phase information is irrelevant.

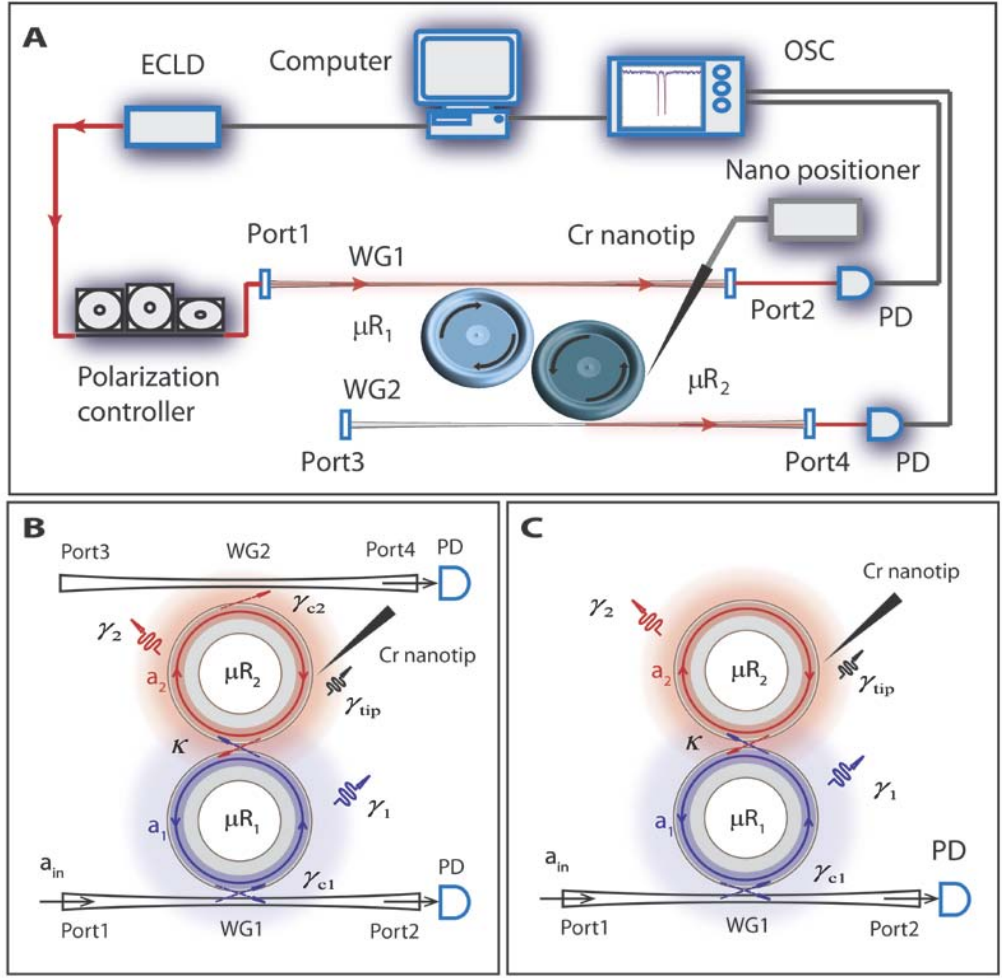

*FIG. S1. Illustration of the experimental setup and the configurations used in the theoretical models and experiments.* *(A)* Experimental setup. PD: photodetector. OSC: oscilloscope. ECLD: external cavity laser diode. $\mu R_1$ and $\mu R_2$: silica microtoroid resonators *(32)*, Cr: Chromium, WG1 and WG2: fiber-taper couplers to couple light in and out of the WGM of the resonators. *(B)* Configuration used in the experiments of the data given in **Fig. 3A-3C**. *(C)* Configuration (with $\gamma_{c2}=0$) used to show the EP phase transition presented in **Fig. 2**, thermal response presented in **Fig. 3D** and the Raman laser spectra given in **Fig. 4** of the main text. See the text for the definitions of the parameters in (B) and (C).

A negative prefactor multiplying $a_1$ ($a_2$) in a first order differential equation for exactly this amplitude $a_1$ ($a_2$) necessarily leads to attenuation, however, this is not the case for a negative prefactor multiplying the amplitude $a_{in}$ for which the phase is not fixed. Thus, the sign of the in-coupling terms can be taken to be positive or negative without affecting the outcome *(31)*.

Rewriting (S.1) in the matrix form, we obtain

$$\frac{d}{dt}\begin{pmatrix} a_1 \\ a_2 \end{pmatrix} = -i \underbrace{\begin{pmatrix} \omega_1 - i\frac{\gamma_1 + \gamma_{c1}}{2} & \kappa \\ \kappa & \omega_2 - i\frac{\gamma_2 + \gamma_{c2} + \gamma_{tip}}{2} \end{pmatrix}}_{M} \begin{pmatrix} a_1 \\ a_2 \end{pmatrix} - \sqrt{\gamma_{c1}}\begin{pmatrix} a_{in} \\ 0 \end{pmatrix}$$

$$= -iM\begin{pmatrix} a_1 \\ a_2 \end{pmatrix} - \sqrt{\gamma_{c1}}\begin{pmatrix} a_{in} \\ 0 \end{pmatrix}$$

(S.2)

Then the characteristic equation and the eigenfrequencies of the system can be found from $|\omega I - M| = 0$. Consequently, the eigenfrequencies of the supermodes formed due to the coupling of the resonators are obtained as

$$\omega_\pm = \frac{1}{2}(\omega_1 + \omega_2) - i\chi \pm \frac{1}{2}\sqrt{4\kappa^2 + \left[(\omega_1 - \omega_2) + i2\Gamma\right]^2}$$

(S.3)

which are complex with the real and imaginary parts defined respectively as $\omega'_\mp$ and $\omega''_\mp$. In (S.3), we have $\chi = (\gamma'_1 + \gamma'_2 + \gamma_{tip})/4$, $\Gamma = (\gamma'_2 + \gamma_{tip} - \gamma'_1)/4$, $\gamma'_1 = \gamma_1 + \gamma_{c1}$ and $\gamma'_2 = \gamma_2 + \gamma_{c2}$, and the expression within the square-root quantify the interplay of the resonator-resonator coupling strength and the loss-contrast of the resonators. Since in the experiments we tune the resonance frequencies of the resonators to be degenerate, we set $\omega_1 = \omega_2 = \omega_0$. In this case Eq. (S.3) reduces to

$$\omega_\pm = \omega_0 - i\chi \pm \beta,$$

(S.4)

where $\beta = \sqrt{\kappa^2 - \Gamma^2}$ with $\beta$ quantifying the effect of the inter-resonator coupling strength $\kappa$ and of the loss-contrast $\Gamma$ between the resonators. Note that $\beta$ can be real or imaginary depending on whether the square of the inter-resonator coupling strength $\kappa$ is greater or smaller than the square of the loss-contrast $\Gamma$.

Substituting $a_k = A_k e^{-i\omega t}$ and $\dfrac{da_k}{dt} = -i\omega A_k e^{-i\omega t} + \dfrac{dA_k}{dt} e^{-i\omega t}$ in Eq. (S1), we find the rate equations

$$\frac{dA_1}{dt} = (i\Delta_1 - \frac{\gamma_1 + \gamma_{c1}}{2})A_1 - i\kappa A_2 - \sqrt{\gamma_{c1}} A_{in}$$

$$\frac{dA_2}{dt} = (i\Delta_2 - \frac{\gamma_2 + \gamma_{c2} + \gamma_{tip}}{2})A_2 - i\kappa A_1$$

(S.5)

where $\Delta_1 = \omega - \omega_1$ and $\Delta_2 = \omega - \omega_2$ are the detuning between the resonance frequencies and the frequency of the input laser light. The input-output relations then become $A_{out2} = A_{in} + \sqrt{\gamma_{c1}} A_1$ and $A_{out4} = \sqrt{\gamma_{c2}} A_2$. Solving Eq. (S.5) at steady state, we obtain the intracavity fields $A_1$ and $A_2$ as

$$A_1 = -\frac{2\sqrt{\gamma_{c1}}\left(\gamma_2' + \gamma_{tip} - i2\Delta_2\right)}{4\kappa^2 + \left(\gamma_1' - i2\Delta_1\right)\left(\gamma_2' + \gamma_{tip} - i2\Delta_2\right)} A_{in}$$

$$A_2 = i\frac{4\kappa\sqrt{\gamma_{c1}}}{4\kappa^2 + \left(\gamma_1' - i2\Delta_1\right)\left(\gamma_2' + \gamma_{tip} - i2\Delta_2\right)} A_{in}$$

(S.6)

The amplitude transmission coefficient $t_{1\to 2}$ of the system from the input port 1 to the output port 2 is given by $t_{1\to 2} = A_{out2}/A_{in}$ which, together with the input-output relation $A_{out2} = A_{in} + \sqrt{\gamma_{c1}} A_1$, leads to $|A_1|^2 = |A_{in}|^2 |t_{1\to 2} - 1|^2 / \gamma_{c1}$. Defining $t_{1\to 2} = |t_{1\to 2}| e^{i\theta}$, we arrive at

$$I_1 = |A_1|^2 = \frac{\left(T_{1\to 2} - 2\sqrt{T_{1\to 2}}\cos\theta + 1\right)}{\gamma_{c1}} |A_{in}|^2$$

(S.7)

where we have defined $T_{1\to 2} = |t_{1\to 2}|^2$. From the input-output relation $A_{out4} = \sqrt{\gamma_{c2}} A_2$, we find $|A_2|^2 = |A_{in}|^2 |t_{1\to 4}|^2 / \gamma_{c2}$ leading to

$$I_2 = |A_2|^2 = \frac{T_{1\to 4}}{\gamma_{c2}}|A_{in}|^2, \tag{S.8}$$

where we have used $t_{1\to 4} = A_{out4}/A_{in}$ and $T_{1\to 4} = |t_{1\to 4}|^2$. One can easily show that the transmittance $T_{1\to 2}$ of the system from the input port 1 to the output port 2 is

$$T_{1\to 2} = \left|\frac{A_{out2}}{A_{in1}}\right|^2 = \left|1 - \frac{2\gamma_{c1}(\gamma_2' + \gamma_{tip} - i2\Delta)}{4\kappa^2 + (\gamma_1' - i2\Delta)(\gamma_2' + \gamma_{tip} - i2\Delta)}\right|^2, \tag{S.9}$$

and similarly the transmittance $T_{1\to 4}$ from the input port 1 to the output port 4 becomes

$$T_{1\to 4} = \left|\frac{A_{out4}}{A_{in1}}\right|^2 = \left|\frac{4\kappa\sqrt{\gamma_{c1}\gamma_{c2}}}{4\kappa^2 + (\gamma_1' - i2\Delta_1)(\gamma_2' + \gamma_{tip} - i2\Delta_2)}\right|^2 \tag{S.10}$$

## S2. Estimating system parameters, eigenfrequencies and intracavity field intensities from experimentally-obtained transmission spectra.

In our experiments, we know the exact values of the resonance frequencies $\omega_0$ of the solitary resonators and their intrinsic losses $\gamma_1$ and $\gamma_2$ (i.e., not including the coupling losses). The other parameters such as $\kappa$, $\gamma_{tip}$, $\gamma_{c1}$ and $\gamma_{c2}$ either are not directly accessible in the experiments (e.g., $\kappa$) or we do not known their exact values (e.g., $\gamma_{c1}$ and $\gamma_{c2}$). In order to estimate these parameters and extract eigenfrequencies as well as the intracavity field intensities, we performed curve-fitting of the analytical expressions from the theoretical model to the experimentally-obtained transmission spectra.

The first set of experiments performed in this study was designed to probe the evolution of the eigenfrequencies of the system. For this purpose, the experimental setup was configured as in **Fig. S1(C)**. Note that in this case $\gamma_{c2} = 0$. Therefore, we had access only to the experimentally-obtained transmission spectra $T_{1\to 2}$. The eigenfrequencies shown in **Fig. 2** of the main text were then estimated by fitting the analytical expression $T_{1\to 2}$ given in Eq. (S.9) to the experimentally-obtained transmission spectra $T_{1\to 2}$. Experiments were performed for different $\kappa$ and $\gamma_{tip}$. A detailed description of the process and its validation were given in detail in the work by Peng *et al.* (18).

In the second set of experiments, we aimed to estimate the intensities of the intracavity fields of the resonators. For this experiment, our setup was configured as in **Fig. S1(B)**. Thus, we had access to the experimentally-obtained transmission spectra $T_{1\to 2}$ and $T_{1\to 4}$. In order to extract the values of the unknown parameters, we performed simultaneous curve fitting of the analytical expressions $T_{1\to 2}$ and $T_{1\to 4}$ given in Eqs. (S.9) and (S.10) to the experimentally-obtained spectra. The curve-fitting provided the estimated values of the parameters of interest with a confidence larger than 0.95. We used the estimated values of the parameters in Eq. (S.6) to calculate the intracavity fields and their intensities, as well as the eigenmode fields.

Equation (S.8) implies that the intracavity field intensity $I_2$ of μR$_2$ can be estimated directly from the measured transmission spectra $T_{1\to 4}$ if the resonator-fiber taper coupling strength $\gamma_{c2}$ is known. On the other hand, Eq. (S.7) implies that in order to estimate the intracavity field intensity $I_1$ of μR$_1$, the measured transmission spectra $T_{1\to 2}$, the resonator-fiber taper coupling strength $\gamma_{c1}$, and the phase $\theta$ of the amplitude transmission coefficient $t_{1\to 2}$ are needed. The dependence on the phase is due to the fact that the measured transmission in port 2 originates from the interference of the part of the input field that directly goes to port 2 (ballistic field) and the part which first couples into the coupled-resonator system and then couples out to the fiber taper to travel to port 2. In the intensity measurement with the photodetector to obtain the transmission spectra, this phase information is lost. From Eq. (S.7), we see that $I_1$ can take any value between its maximum value $\max(I_1) = \left(\sqrt{T_{1\to 2}} + 1\right)^2 |A_{in}|^2 / \gamma_{c1}$ at $\theta = \pi$ and its minimum value $\min(I_1) = \left(\sqrt{T_{1\to 2}} - 1\right)^2 |A_{in}|^2 / \gamma_{c1}$ at $\theta = 0$. Using Eq. (S9), we find that $\theta = 0$ when $\Delta = 0$ (zero detuning). This implies that in order to estimate $I_1$ at the frequency $\omega_0$, we can take $\theta = 0$. This phase can be derived from Eq. (S.9) by grouping it into its real and imaginary parts as $t_{1\to 2} = \text{Re}[t_{1\to 2}] + i\,\text{Im}[t_{1\to 2}] = |t_{1\to 2}|e^{i\theta}$. Then the phase $\theta$ is given by $\theta = \text{atan}\left(\text{Im}[t_{1\to 2}]/\text{Re}[t_{1\to 2}]\right)$. In our experiments, $\theta$ changes only within a very small region around zero degrees and thus $0.993 \leq \cos\theta \leq 1$. Therefore, we can safely take $\cos\theta = 1$ in the process of estimating $I_1$ from the measured $T_{1\to 2}$. At $\theta = 0$, the estimated value of $I_1$ will correspond to $\min(I_1)$. The real intracavity field intensity of µR$_1$ may be larger than the estimated value.

In short, we perform curve-fitting of the analytical expressions of transmission spectra, obtained from our theoretical model, to the experimentally-obtained transmission spectra to obtain the values of the system parameters. The estimated values of these parameters were within the expected range in our system. For example, by performing many measurements using a WG1-µR$_1$ (and WG2-µR$_2$) system alone, we estimated possible values of $\gamma_{c1}$ (and $\gamma_{c2}$), and performed the curve-fitting such that the estimated values are within the experimentally possible ranges. The estimated values were then used with the analytical expressions given in Eqs. (S.3)-(S.10) to calculate eigenfrequencies, intracavity fields and their intensities.

### S3. Transmission spectra of the coupled-resonators system at different coupling regimes.

Here we give theoretical results using Eqs. (S.9) and (S.10) of **Sec. S1** and typical transmission spectra obtained in the experiments at different coupling regimes. We consider the configuration shown in **Fig. S1(B)**. Note that this configuration was used to obtain the results depicted in **Figs. 3(A)-3(C)** of the main text. Here each of the resonators µR$_1$ and µR$_2$ was coupled to a different fiber-taper. Thus both of the resonators experienced coupling losses in addition to their intrinsic loss. Moreover, the second resonator µR$_2$ experienced the additional loss $\gamma_{\text{tip}}$. We tested two different cases by choosing different mode pairs in the resonators. *Case 1:* The mode chosen in µR$_1$ had higher loss than the mode in µR$_2$, that is $\gamma_1 + \gamma_{c1} > \gamma_2 + \gamma_{c2}$. *Case 2:* The mode chosen in µR$_2$ had higher loss than the mode in µR$_1$, that is $\gamma_1 + \gamma_{c1} < \gamma_2 + \gamma_{c2}$. In both cases, the additional loss $\gamma_{\text{tip}}$ was introduced to the mode in the second resonator µR$_2$. Therefore, in the first case additional loss $\gamma_{\text{tip}}$ was induced in the mode with less loss whereas in the second case $\gamma_{\text{tip}}$ was induced in the mode with higher loss.

Although general features of the transmission spectra for *Case 1* and *Case 2* are the same, there are differences in their response to varying $\gamma_{tip}$. First, the amount of additional loss $\gamma_{tip}$ required to bring the system to the exceptional point (EP), where the split modes coalesce, is higher for *Case 1* than for *Case 2*. Depending on the initial loss contrast, even a small amount of $\gamma_{tip}$ may complete the transition from the strong- to weak-coupling regime through the EP. Second, in the weak-coupling regime the depth of the resonance for *Case 2* is deeper than that for *Case 1*, because in this case the resonators cannot exchange energy easily and the transmission $T_{1\to 2}$ is effectively determined by the loss of the first cavity which has less loss. In *Case 1*, the WG1-resonator system has already a large amount of loss, therefore moving the system into the deep weak-coupling regime does not affect the WG1-resonator coupling condition; that is, the system stays further away from the critical coupling (that is the waveguide-resonator coupling loss equals to the sum of all other losses such as absorption, scattering, and radiation losses) when compared to the *Case 2*.

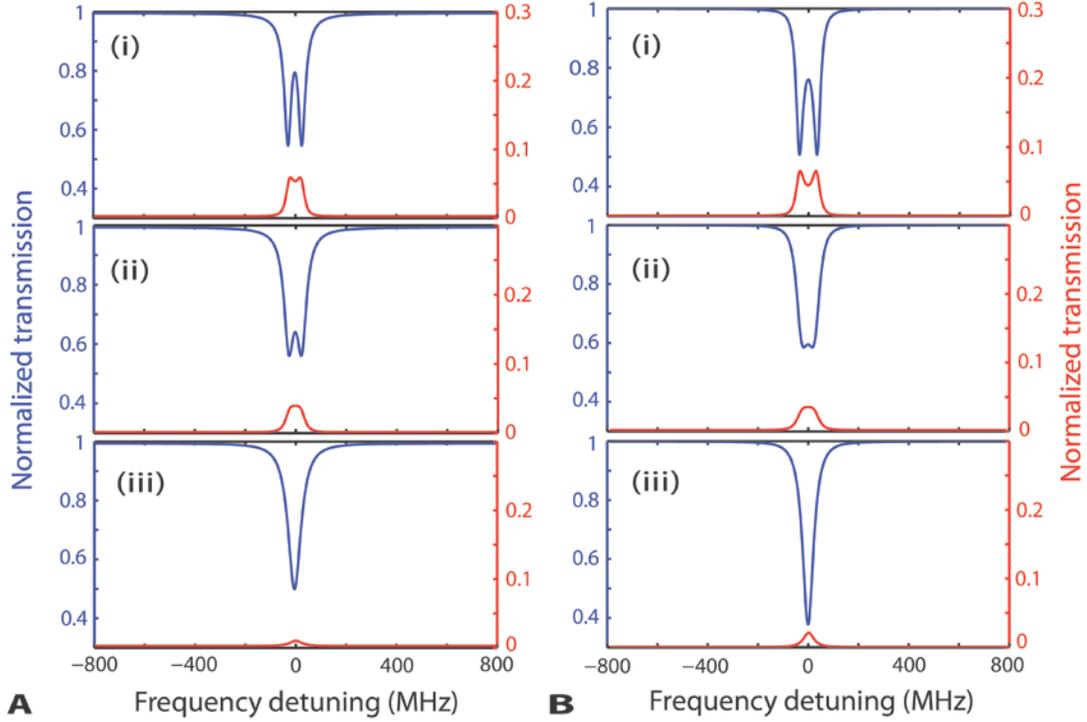

*Fig. S2. Theoretical normalized transmission spectra $T_{1\to 2}$ (blue) and $T_{1\to 4}$ (red) for Case 1 (A) and Case 2 (B). Additional loss induced in the second resonator is increased via $\gamma_{tip}$ from top to bottom moving the system from strong (i) to moderate (ii) and to weak (iii) coupling regimes through the exceptional point. Case 1: Initial loss of the first resonator is higher than that of the second resonator where the additional loss $\gamma_{tip}$ is introduced. Case 2: Initial loss of the first resonator is lower than that of the second resonator where the additional loss $\gamma_{tip}$ is introduced.*

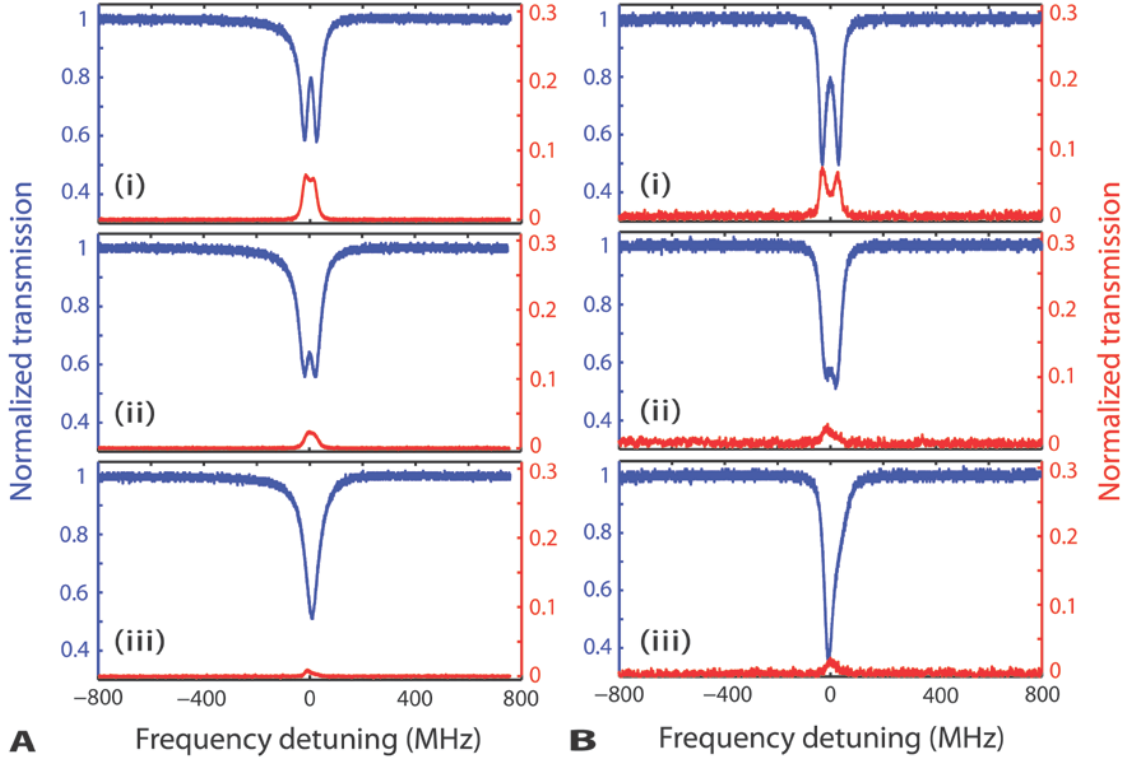

*Fig. S3. Experimentally-obtained normalized transmission spectra $T_{1\to 2}$ (blue) and $T_{1\to 4}$ (red) for Case 1 (A) and Case 2(B). Additional loss induced in the second resonator is increased via $\gamma_{tip}$ from top to bottom moving the system from strong (i) to moderate (ii) and to weak (iii) coupling regimes through the exceptional point.*

In **Fig. S2**, we give the transmission spectra obtained from Eqs. (S.9) and (S.10) for Case 1 and Case 2. The parameter values used in **Fig. S2** are typical values observed in our experiments. **Figure S3** depicts the experimentally-obtained transmission spectra for Case 1 and Case 2. In both of the cases, the theoretical and experimental transmission spectra agree very well, and the differences between the Case 1 and Case 2 mentioned in the previous paragraph are clearly seen. For example, in the spectra labeled as (iii) which correspond to the weak-coupling regime (large $\gamma_{tip}$) in **Fig. S2** and **Fig. S3**, we see that the theoretical and experimental resonances are deeper for Case 2 than in Case 1. Similarly, in (ii) we see that although the same amount of $\gamma_{tip}$ is used, the spectra of Case 2 are closer to the EP than that of Case 1 (i.e., the splitting in Case 1 is more resolved than that in Case 2, implying that Case 2 is closer to the EP than Case 1).

## S4. Evolution of the eigenfrequencies of the coupled-resonator system as a function of the coupling strength $\kappa$ and the loss $\gamma_{tip}$.

In the experiments, we tune the resonances of the resonators to be degenerate (i.e., $\omega_1 = \omega_2 = \omega_0$). Thus we can use Eq. (S.4) to understand the movement of eigenfrequencies as a function of the additional loss $\gamma_{tip}$ and $\kappa$. From Eq. (S.4), we see that the spectral distance between the eigenfrequencies of the system is given by $\delta = \omega_+ - \omega_- = 2\beta$. It is clear that when the resonators are placed far from each other such that they cannot exchange energy (i.e., *no inter-resonator coupling* $\kappa = 0$), we have two solitary resonators with resonances having the same resonance frequencies $\omega_0$ (i.e., no mode-splitting) but different linewidths quantified by the total losses experienced by each of the resonators: $\gamma_1 + \gamma_{c1}$ for µR$_1$ and $\gamma_2 + \gamma_{c2} + \gamma_{tip}$ for µR$_2$.

When the two resonators are placed in each other's vicinity, the resonant modes overlap, leading to a finite coupling strength. For a fixed $\gamma_{tip}$, the transition between the real and imaginary $\beta$ occurs at the threshold inter-resonator coupling strength $\kappa_{EP} = |\Gamma|$ that leads to $\beta = 0$ (*Exceptional point: EP*). Similarly, when the inter-resonator coupling strength is fixed at $\kappa$, this transition takes place at the threshold $\gamma_{tip}$ given by

$$\gamma_{tip}^{EP} = (\gamma_1' - \gamma_2') \mp 4\kappa. \tag{S.11}$$

Note that $\gamma_{tip}^{EP} > 0$ corresponds to introducing additional loss whereas $\gamma_{tip}^{EP} < 0$ corresponds to introducing gain. At the transition point, there is zero spectral distance between the supermodes (i.e., coalescence in real parts), and a linewidth equal to the average of the linewidths of the solitary resonators. Note that the EP takes place at different $\gamma_{tip}$ for different $\kappa$ and vice versa.

The *Strong-coupling regime* is quantified by $\kappa > |\Gamma|$ and hence by real $\beta$. As a result, the supermodes have different resonance frequencies, spectrally separated from each other by $2\beta$. We referred to this as mode splitting. The supermodes have the same linewidths quantified by $\chi$ which is determined by the total losses experienced by the two resonators.

The *weak-coupling regime* is quantified by $\kappa < |\Gamma|$ and hence by an imaginary $\beta$. Consequently, the two supermodes have the same resonance frequency but different linewidths. We plotted Eqs. (S.3) and (S.4) as a function of $\gamma_{tip}$ in **Fig. S4** which show the movement of the eigenfrequencies in the complex plane when the coupled resonance modes in the resonators are initially zero-detuned [**Fig. S4(A)**] and non-zero detuned [**Fig. S4(B)**].

The color bar in **Fig. S4** has the same meaning as the y-axis of the figure, that is, it represents the value of the imaginary parts of the supermodes. The upper bound for the values of the imaginary parts of the eigenvalues of the supermodes is determined by the decay rate of the first cavity $\gamma'_1/2$. When $\gamma_{tip}$ is introduced to μR$_2$ and gradually increased, both of the supermodes are affected in the same way and imaginary parts of their eigenvalues decrease (become more negative). After a critical value of the additional loss, one of the eigenmodes starts to become more localized in µR$_1$ whereas the other mode becomes localized in µR$_2$. The mode in µR$_2$ feels the effect of the additional loss more and more strongly such that the imaginary part of its eigenvalue becomes more negative. At the same time, since the system moves into the weak-coupling regime and the other mode is more localized in µR$_1$, it feels the effect of the additional loss less and less strongly. As a result, the imaginary part of its eigenvalue first decreases (becomes more negative) and then starts increasing (becomes less negative) as $\gamma_{tip}$ is increased. However, the value of the imaginary part of its eigenvalue cannot go below (become less negative than) its initial value before the additional loss $\gamma_{tip}$ is introduced to µR$_2$, because although being very small, the mode still sees the additional loss. Since our system is a passive system for the mode at 1550 nm band, no additional energy is supplied to compensate the system loss, such that the system's supermodes can never have a linewidth smaller than their initial values, that is their imaginary parts cannot be larger than their initial values at $\gamma_{tip} = 0$.

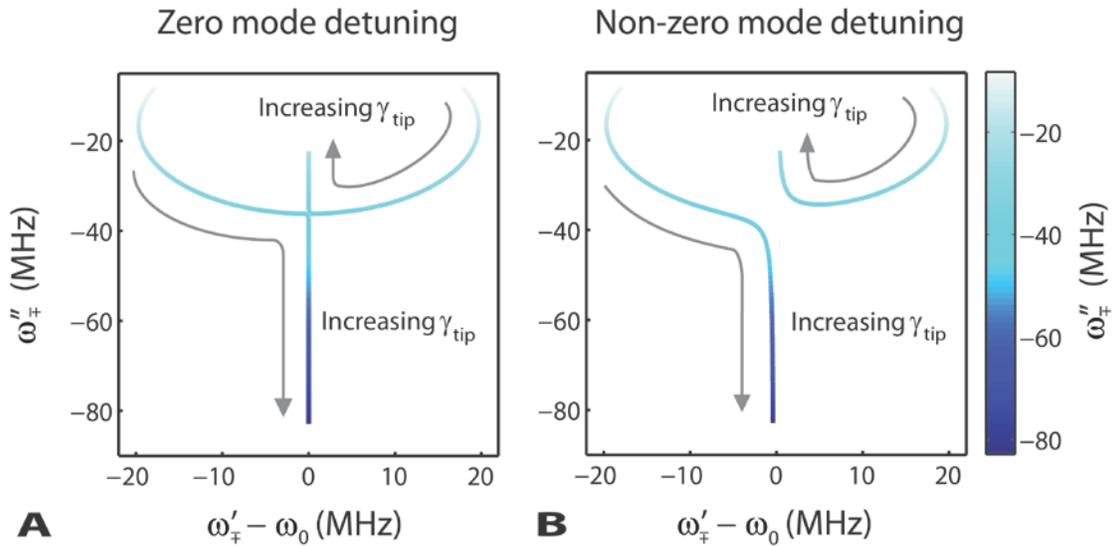

*Fig. S4. Evolution of the real $\omega'_\mp$ and imaginary $\omega''_\mp$ parts of the eigenfrequencies of the coupled-resonators system as a function of additionally induced loss $\gamma_{tip}$. (A) Solitary resonators have the same resonance frequencies $\omega_0$: zero-detuning. (B) Solitary resonators have different resonance frequencies: non-zero detuning. An avoided crossing is clearly seen.*

Our scheme allows steering the system via both the cavity-cavity coupling strength $\kappa$ and the loss $\gamma_{tip}$ using the same coupled resonators. Benefiting from this feature, we performed experiments at varying coupling strengths and losses to understand the parametric dependence of the eigenfrequency movement in the complex plane and to elucidate the trade-off between $\gamma_{tip}$ and $\kappa$ when the system is steered through the EP. By tuning $\kappa$ and $\gamma_{tip}$, we drove the coupled-resonator system into different regimes and probed the evolution of the complex eigenfrequencies. We performed experiments using the configuration shown in **Fig. S1(C)**. Note that this configuration was used to obtain the results depicted in **Fig. 2** of the main text. Here only the first resonator µR$_1$ was coupled to a fiber-taper and experienced the coupling loss $\gamma_{c1}$ in addition to its intrinsic loss $\gamma_1$. The second resonator µR$_2$, on the other hand, was not coupled to a fiber-taper and hence did not experience the coupling loss $\gamma_{c2}$ (that is, $\gamma_{c2} = 0$).

Since the nanotip that was used to induce the additional loss $\gamma_{tip}$ was introduced into the mode volume of µR$_2$, this resonator experienced the loss $\gamma_{tip}$ in addition to its intrinsic loss $\gamma_2$ but without any coupling loss ($\gamma_{c2} = 0$). The results of this experiment have been summarized in **Fig. 2** and the related parts in the main text. Here in **Fig. S5** and **Fig. S6**, we provide additional results which depict the difference between the real [**Figs. S5(A), S6(A)**] and imaginary [**Figs. S5(B), S6(B)**] parts of the eigenfrequencies as a function of the induced loss $\gamma_{tip}$ and the inter-resonator coupling strength $\kappa$. It is seen that the coalescence of eigenfrequencies (EP) occurs at different $\gamma_{tip}$ for different $\kappa$: The EP transition occurred at higher $\gamma_{tip}$ for stronger $\kappa$ (**Figs. S5** and **S6**). After the EP, the imaginary parts of the eigenfrequencies bifurcate with different slopes. As a result, one of the supermodes experiences significantly higher loss than the other [**Figs. S5(B)** and **S6(B)**].

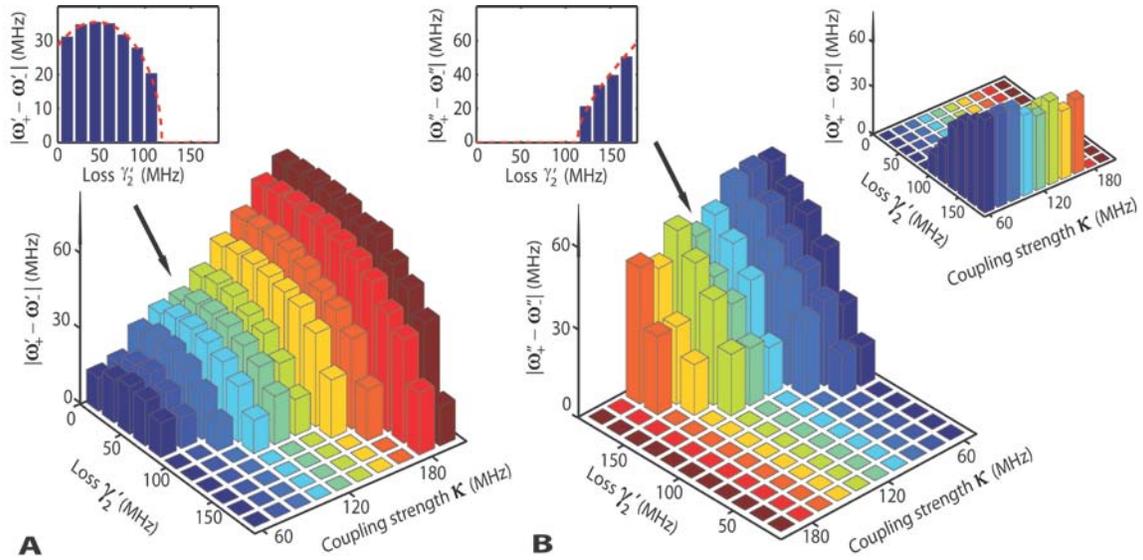

*Fig. S5. Evolution of the real and imaginary parts of the eigenfrequencies of the supermodes as a function of the loss in the second resonator $\gamma_2'$ and coupling strength $\kappa$. (A) Difference of the real parts of the eigenfrequencies: $|\omega_+' - \omega_-'|$. (B) Difference in the imaginary parts of the eigenfrequencies: $|\omega_+'' - \omega_-''|$. The loss of the second resonator was varied using the chromium nanotip which introduced the additional loss of $\gamma_{tip}$. As shown in the top left inset $|\omega_+' - \omega_-'|$ first increased and then decreased as $\gamma_{tip}$ was increased. This indicated that $\gamma_1 + \gamma_c > \gamma_2$ when $\gamma_{tip} = 0$ (see main text). The top right inset shows the same plot as in the main figure, but with the oriented axes as in (A).*

We extracted the eigenfrequencies from the experimentally obtained transmission spectra by curve-fitting the theoretical $T_{1 \to 2}$ given in Eq. (S.9). In this process, we used the measured parameters $\gamma_1$, $\gamma_2$ and $\gamma_{c1}$ and set $\omega_1 = \omega_2 = \omega_0$ ($\Delta_1 = \Delta_2 = \Delta = \omega - \omega_0$), $\gamma_2' = \gamma_2$ (i.e., $\gamma_{c2} = 0$). Other parameters were left as free parameters for curve-fitting. A detailed description of the process was given in **Sec. S2** and in Ref. *(18)*.

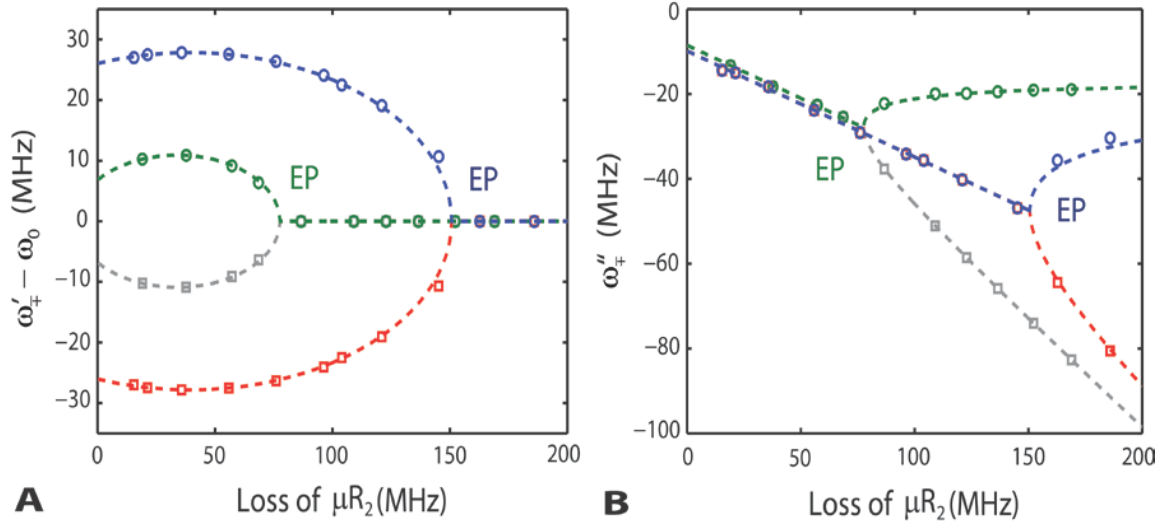

*Fig. S6. Evolution of the real and imaginary parts of the eigenfrequencies of the supermodes as a function of the loss in the second resonator $\gamma_2'$ at different inter-resonator coupling strengths $\kappa$. (A) Real and (B) imaginary parts of the eigenfrequencies are shown as a function of the loss in the second resonator $\gamma_{tip}$ at two different coupling strengths $\kappa$. The stronger the coupling $\kappa$, the larger the loss $\gamma_{tip}$ required in the second resonator to observe an exceptional point (EP). Data represented by open squares and circles denote the experimentally obtained eigenfrequencies. Dashed red and blue lines correspond to the best fit to the experimental data using the theoretical model (Sec. S1).*

S14

## S5. Intracavity field intensities of the coupled resonator system.

We have calculated the intracavity field intensities of the resonators at the eigenfrequencies $\omega_{\pm}$ and at the resonance frequency $\omega_0$ as a function of $\gamma_{tip}$ using Eq. (S.6) and plotted them in **Figs. S7-S9**. Note that we chose to look at the frequency $\omega_0$ in addition to $\omega_{\pm}$ because the two eigenfrequencies $\omega_+$ and $\omega_-$ coalesce at $\omega_0$ at the EP and retain this value after the EP throughout the weak-coupling regime. It is clearly seen that when a specific eigenfrequency is excited, the evolution of the intracavity field intensities $I_1 = |A_1|^2$ and $I_2 = |A_2|^2$ as a function of $\gamma_{tip}$ differs significantly. As expected, intracavity field intensities at $\omega_+$ and $\omega_-$ evolve similarly but different from what is observed at $\omega_0$.

At $\omega_{\pm}$, the intracavity field intensities are close to each other when $\gamma_{tip}$ is zero, an indication that initially the supermodes are distributed almost equally between the resonators. As $\gamma_{tip}$ is increased, both $I_1$ and $I_2$ first decrease with increasing difference in their intensities until they reach their local minimum at $\gamma_{tip} = \gamma_{tip}^{min}$. The difference in their rates of decrease is due to the fact that the total loss of the second resonator is continuously increased by $\gamma_{tip}$. Note that in this region $\gamma_{tip}$ is not strong enough to significantly affect the distribution of the supermodes. As $\gamma_{tip}$ is increased further to bring the system closer to the EP, the intracavity field intensities starts increasing. Here, the increase of $I_1$ is significant but that of $I_2$ is very small (the increase in $I_2$ is barely seen: green curves **Fig. S7** and **Fig. S8**). This trend continues until $\gamma_{tip} = \gamma_{tip}^{EP}$ beyond which $I_1$ continues to increase whereas $I_2$ starts to decrease again. This is because after the EP, one of the eigenmodes becomes more localized in the first resonator which has less loss than the second, and the other eigenmode becomes more localized in the second resonator. Consequently, the total field feels less loss compared to the initial point when $\gamma_{tip} = 0$. As a result, the total intensity $I_T = I_1 + I_2$ after the EP (in the weak-coupling regime) is larger than that before the EP despite increasing $\gamma_{tip}$. Note that as $\gamma_{tip}$ continues to increase beyond $\gamma_{tip}^{EP}$, the total intracavity field intensity $I_T$ approaches the intracavity field intensity of the first resonator, while the intracavity field intensity of the second resonator continuously decreases, becoming negligible. This is in contrast to the expectation that the intensity would decrease with increasing loss, and is a direct consequence of the effect of the EP.

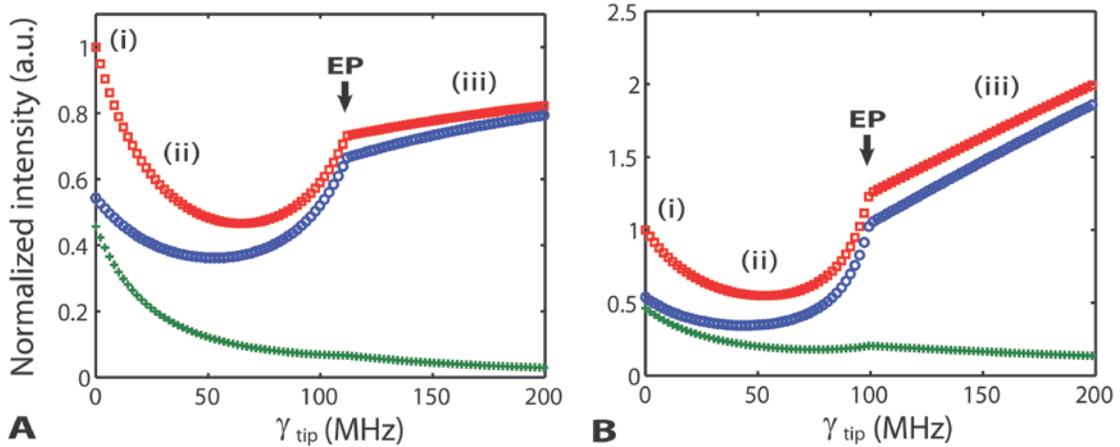

*Fig. S7. Theoretically obtained normalized intracavity field intensities of the coupled resonators at the eigenfrequency $\omega_+$. Blue (Green): Intensity $I_1$ ($I_2$) in the first (second) resonator. Red: Total Intensity $I_T = I_1 + I_2$. Normalization was done with respect to the total intensity at $\gamma_{tip} = 0$. (A) Case 1: Initial loss of the first resonator is higher than that of the second resonator. (B) Case 2: Initial loss of the first resonator is lower than that of the second resonator. (i)-(iii) correspond to coupling regimes as in **Fig. S2** and **Fig. S3**. EP: Exceptional point.*

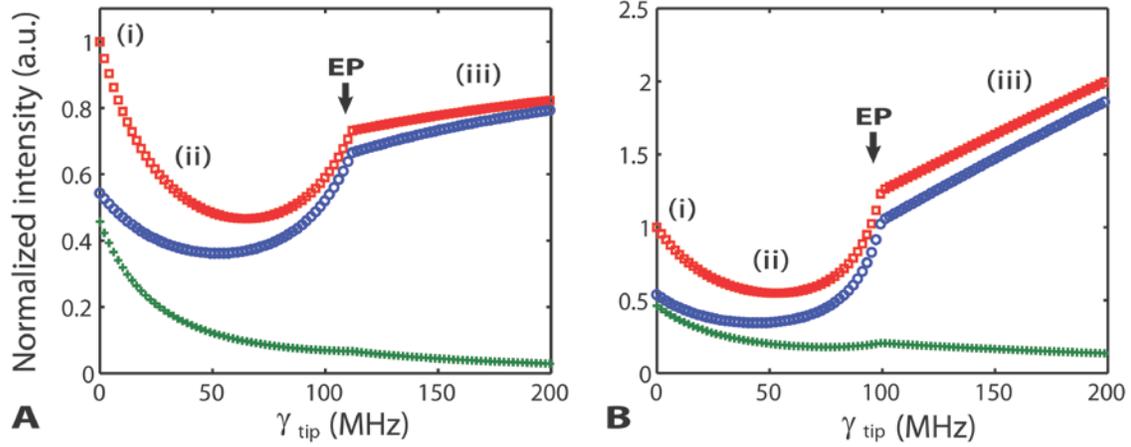

*Fig. S8. Theoretically obtained normalized intracavity field intensities of the coupled resonators at the eigenfrequency $\omega_-$. Color codes are the same as in **Fig. S7**. (**A**) Case 1: Initial loss of the first resonator is higher than that of the second resonator. (**B**) Case 2: Initial loss of the first resonator is lower than that of the second resonator. Evolution of intracavity intensities for $\omega_-$ and $\omega_+$ (**Fig. S7**) are the same. (i)-(iii) correspond to coupling regimes as in **Fig. S7**. EP: Exceptional point.*

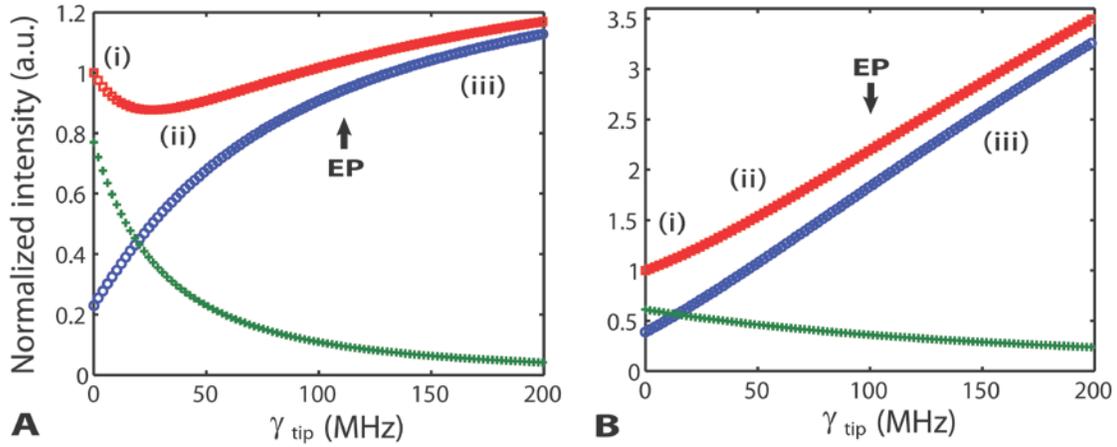

*Fig. S9. Theoretically obtained normalized intracavity field intensities of the coupled resonators at the eigenfrequency $\omega_0$. Color codes are the same as in **Fig. S7** and **Fig. S8**. Note that $\omega_0$ is the eigenfrequency at the exceptional point. (**A**) Case 1. (**B**) Case 2.*

For intracavity field intensities at $\omega_0$ (see **Fig. S9**), we find that when $\gamma_{tip}$ is zero, the field is highly localized in the second resonator ($I_1 < I_2$) for *Case 1* [**Fig. S9(A)**]. For *Case 2*, on the other hand, the field is almost evenly distributed between the resonators [**Fig. S9(B)**]. With increasing $\gamma_{tip}$, $I_2$ decreases while $I_1$ increases. As a result, $I_T$ first decreases reaching a minimum value and then increases. Note that in the large $\gamma_{tip}$ limit, the total intracavity field approaches that of the intracavity field of the first resonator, implying that the field is almost completely localized in the first resonator.

In **Fig. 3A** and **Fig. 3B** of the main text we have given $I_1$, $I_2$ and $I_T$ estimated from the experimentally-obtained transmission spectra $T_{1\to 2}$ and $T_{1\to 4}$ at $\omega_+$ for *Case 1* and *Case 2*, respectively. In **Fig. S10** and **Fig. S11**, we provide the experimentally-obtained $I_1$, $I_2$ and $I_T$ at $\omega_-$ and $\omega_0$, respectively. Note that the intensity versus $\gamma_{tip}$ given in **Fig. 3** of the main text for $\omega_+$ and that given in **Fig. S10** for $\omega_-$ are the same; that is the intracavity field intensities of the resonators at the eigenfrequencies are the same. We obtained good agreement between the data points obtained from the experimentally-obtained transmission spectra and the theoretical expectations.

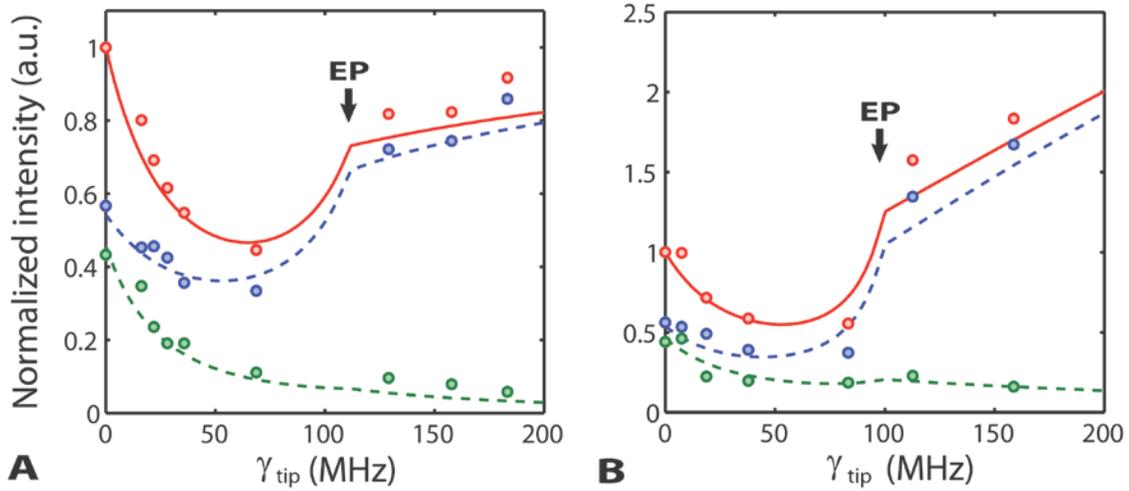

*Fig. S10. Experimentally-obtained intracavity field intensities at the eigenfrequency $\omega_-$. Blue and green curves correspond to intracavity field intensities $I_1$ and $I_2$ of the first and second resonators, respectively. Red curves denote the total intracavity field intensity $I_T = I_1 + I_2$ (sum of green and blue). Circles are experimentally-obtained data whereas the lines are from the theoretical model. EP: Exceptional point. Normalization was done with respect to the total intensity at $\gamma_{tip} = 0$. (A) Case 1: Initial loss of the first resonator is higher than that of the second resonator. (B) Case 2: Initial loss of the first resonator is lower than that of the second resonator.*

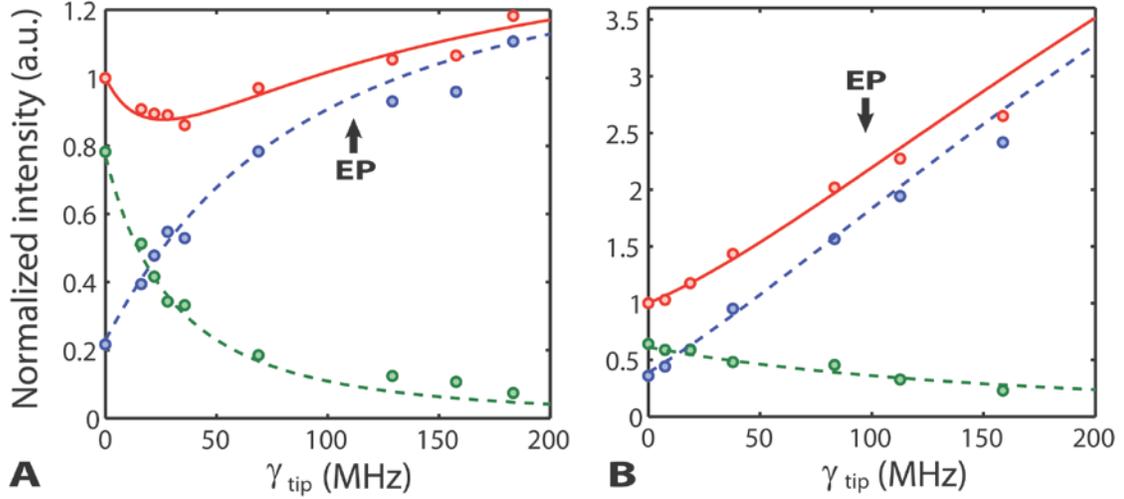

*Fig. S11. Experimentally-obtained intracavity field intensities at the eigenfrequency $\omega_0$. Blue and green curves correspond to intracavity field intensities $I_1$ and $I_2$ of the first and second resonators, respectively. Red curves denote the total intracavity field intensity $I_T = I_1 + I_2$ (sum of green and blue). Circles are experimentally-obtained data whereas the lines are from the theoretical model. EP: Exceptional point. Normalization was done with respect to the total intensity at $\gamma_{tip} = 0$. (A) Case 1: Initial loss of the first resonator is higher than that of the second resonator. (B) Case 2: Initial loss of the first resonator is lower than that of the second resonator.*

In **Fig. S12**, we provide the intracavity field intensities at eigenfrequencies $\omega_\pm$ and $\omega_0$ normalized to the intensities at the EP. It is clearly seen that for $\gamma_{tip} \geq \gamma_{tip}^{EP}$ (i.e., after the EP transition, that is, in the weak coupling regime), the intracavity field intensities at $\omega_\pm$ and $\omega_0$ coincide. This is expected because at the EP, the eigenfrequencies [as given in Eq. (S.3)] of the coupled-resonators system coalesce in their real parts (resonance frequency) and attain the average of the resonance frequencies of the solitary resonators. Since in our theoretical analysis and experiments we have set the resonance frequencies of the solitary resonators the same at $\omega_0$, at the EP the resonance frequencies of the supermodes converge to $\mathrm{Re}[\omega_\mp] = \omega'_+ = \omega'_- = (\omega_0 + \omega_0)/2 = \omega_0$. Therefore the intensities in the weak-coupling regime are the same at $\omega_0$ and at $\omega_\pm$.

In **Fig. 3C**, we gave the experimental data showing that the total intracavity field intensities at $\omega_{\pm}$ and $\omega_0$ coincide at the exceptional point and stay the same after the exceptional point as the additional loss is increased. The data provided in **Fig. 3C** was obtained for the *Case 1*. Here in **Fig. S13**, we provide the experimental data for *Case 1* [**Fig. S13(A)** and also shown in **Fig. 3C**] and *Case 2* [**Fig. S13(B)**] together for comparison purposes. As can be seen, the intracavity field intensities at $\omega_{\pm}$ and $\omega_0$ coincide at the exceptional point for both of the cases.

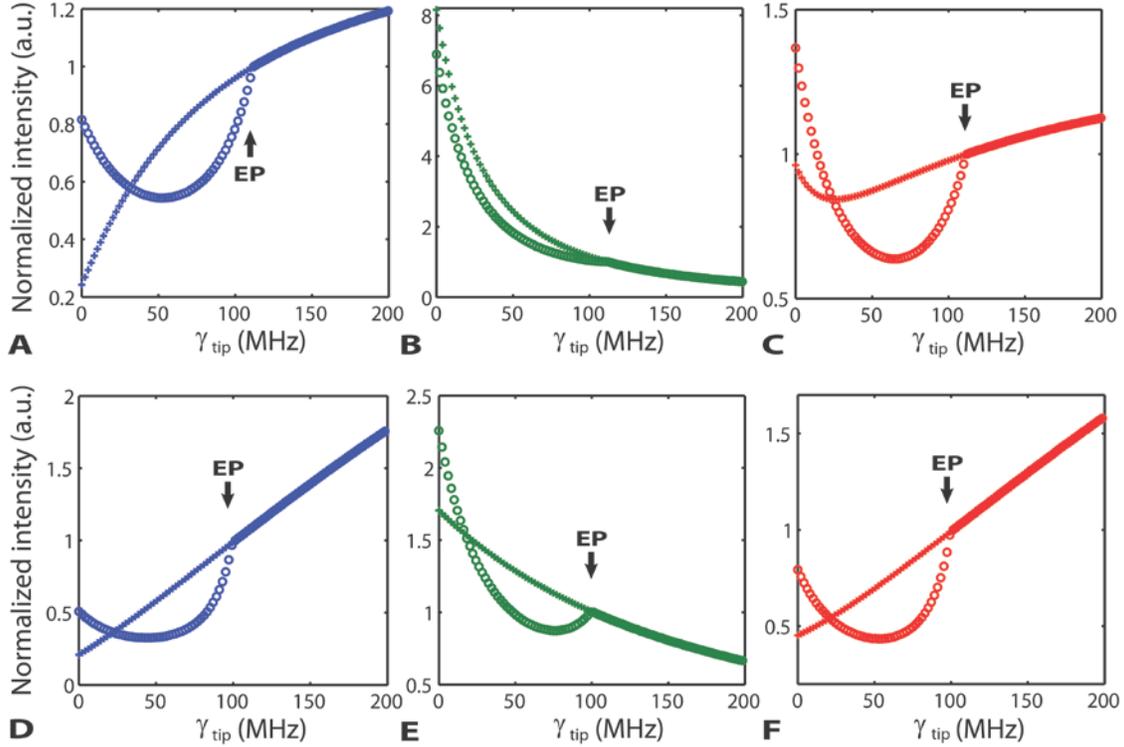

*Fig. S12. Theoretically obtained intracavity field intensities normalized with the intensity at the exceptional point (EP). Intracavity field intensities calculated at eigenfrequencies $\omega_-$ and $\omega_+$ are the same (curves with circles), and they coincide with intracavity field intensities calculated at the eigenfrequency $\omega_0$ of the exceptional point (curves with +). (A)-(C) are obtained for Case 1: Initial loss of the first resonator is higher than that of the second resonator. (D)-(F) are obtained for Case 2: Initial loss of the first resonator is lower than that of the second resonator. Intracavity field intensities $I_1$ (blue curves), $I_2$ (green curves), and $I_T$ (red curves) are respectively given in (A), (B) and (C) for Case 1, and in (D), (E) and (F) for Case 2.*

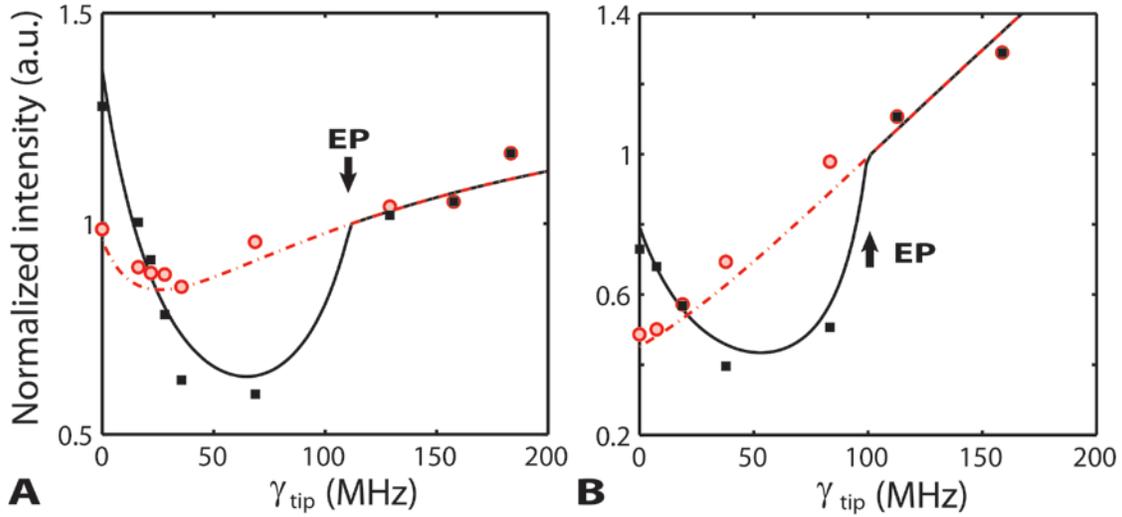

*Fig. S13. Experimentally-obtained total intracavity field intensities normalized with the intensity at the exceptional point (EP). Total intracavity field intensities calculated at eigenfrequencies $\omega_-$ and $\omega_+$ are the same (black), and they coincide with the total intracavity field intensities calculated at the eigenfrequency $\omega_0$ of the exceptional point (red). Black squares and red circles are the data points obtained in the experiments for total intracavity field intensities at $\omega_\mp$ and for $\omega_0$, respectively. Normalization is done with the intensity at the exceptional point. (A) Case 1: Initial loss of the first resonator is higher than that of the second resonator. (B) Case 2: Initial loss of the first resonator is lower than that of the second resonator.*

## S6. Why are the intracavity field intensities nonzero at $\omega_0$ when $\gamma_{tip} = 0$ in Fig. S9 and Figs. S11-S13?

When the system is away from the EP, light can be coupled into the system of coupled-resonators only at its eigenfrequencies $\omega_\mp$. If the splitting (spectral distance) between the eigenfrequencies is so large that there is no spectral overlap between the resonance lineshapes, the eigenfrequencies can be excited individually with a narrow-linewidth laser. That is $2|\omega'_+ - \omega'_-| > (\delta_1 + \delta_2)$ where $\omega'_\mp$ are the resonance center frequencies of the eigenmodes and $\delta_{1,2}$ are their linewidths. Moreover, the cavity field should be zero at any frequency $\omega_m$ other than $\omega_\mp$, assuming that the observed frequency $\omega_m$ does not fall within the linewidths of the resonance lines at $\omega_\mp$, that is

S21

$\omega_m \notin [\omega'_+ - \delta_1/2, \omega'_+ + \delta_1/2]$ and $\omega_m \notin [\omega'_- - \delta_1/2, \omega'_- + \delta_1/2]$. In **Fig. S9**, and **Figs. S11-S13**, on the other hand, we see that when $\gamma_{tip} = 0$ the intracavity field intensities at $\omega_0$ is non-zero. This may seem surprising because when $\gamma_{tip} = 0$, the system is far from the EP where $\omega_+ = \omega_- = \omega_0$. Then one may wonder why the intracavity field intensities of the resonators at $\omega_0$ are nonzero. This originates from the special setting of our system used in measuring the presented data. As we will explain in detail below, this setting violates the above conditions although the system is still in the strong coupling regime ($\kappa > |\Gamma|$) and far from the EP.

In order to avoid nonzero field at $\omega_0$ and satisfy the above conditions, the coupled-resonator system should be in the very strong coupling regime when $\gamma_{tip} = 0$. As $\gamma_{tip}$ increases, the system approaches the EP, that is, the eigenfrequencies $\omega_\mp$ approach each other and the frequency $\omega_0$. The larger $\gamma_{tip}$ becomes, the larger the overlap between the eigenmodes becomes. At the same time, the finite linewidths of the supermode resonances covers $\omega_0$. As a result, the intracavity fields at $\omega_0$ become nonzero and increase as the system approaches the EP. Finally, as we have shown in **Fig. S12** and **Fig. S13** the fields at $\omega_\mp$ and $\omega_0$ coincide in the weak-coupling regime.

In our experiments, the amount of additional loss introduced to the second resonator is, however, limited by the absorption in the chromium-coated nanotip and its overlap with the evanescent field of the second resonator. When we drove the system in the very strong coupling regime, we observed that when $\gamma_{tip} = 0$, the field at $\omega_0$ was zero as expected. However, the amount of loss provided by the nanotip was not enough to move the system from this very strong coupling regime into the weak-coupling regime. Therefore, we set the working regime of the system such that it was still in the strong coupling regime ($\kappa > |\Gamma|$) and far from the EP but within the range that the available additional loss $\gamma_{tip}$ could drive the system into the weak-coupling regime.

In **Fig. S14(A)**, we give the experimentally obtained transmission spectra in the very strong coupling regime when $\gamma_{tip} = 0$. Here, the splitting is large and the resonances at $\omega_\mp$ are well-resolved. The normalized transmission spectra $T_{1\to 2}$ is unity everywhere except in the vicinity of the resonance lineshapes. Therefore, we have $T_{1\to 2} \sim 1$ at $\omega_0$ which is at mid-way between the

S22

two resonances implying that there is no coupling into the resonators at frequency $\omega_0$. Similarly in the normalized transmission spectra $T_{1\to 4}$, the two resonance peaks are clearly seen and $T_{1\to 4} \sim 0$ at frequency $\omega_0$ confirming that light is coupled into the system only at frequencies within the linewidths of the resonance modes. In the transmission spectra given in **Fig. S2**, **Fig. S3** and **Fig. 2**, which formed the basis of the data presented in our experiments, it is clearly seen that $T_{1\to 2}$ is well-below one at $\omega_0$ and $T_{1\to 4}$ is well-above the zero level. This implies the leakage of light power from the eigenmodes into $\omega_0$, since now the resonance lineshapes have nonzero overlap with $\omega_0$.

In **Fig. S14(B)**, we provide the intracavity field intensities at $\omega_\mp$ and $\omega_0$ as a function of $\gamma_{tip}$ when the system is in the very strong coupling regime at $\gamma_{tip} = 0$. We also provide experimental data, which show that when $\gamma_{tip} = 0$ the intracavity field intensity at $\omega_0$ is zero. Starting with the experimentally obtained system parameters $\kappa$, $\gamma_1$, $\gamma_2$, and $\omega_0$ and varying $\gamma_{tip}$, we obtained the theoretical curve in **Fig. S14(B)** and its inset which shows $\gamma_{tip}^{EP} \sim 900$ MHz which was beyond the attainable $\gamma_{tip}$ with our nanotip under this experimental condition. Therefore, we had to move the system closer to the EP but still in the strong coupling regime when $\gamma_{tip} = 0$. This allowed us to move the system from the strong- to the weak-coupling regime with the nanotips we had. As a result of this, we obtained the transmission spectra in **Fig. S2**, **Fig. S3** and **Fig. 2** with resonances of the supermodes closer to each other and non-zero intracavity field at $\omega_0$ when $\gamma_{tip} = 0$. However, before moving the system from its strong coupling-regime closer to EP, we measured the system parameters to estimate the intracavity field intensity at $\omega_0$ when $\gamma_{tip} = 0$ and when $\gamma_{tip} \sim 30$ MHz and depicted them in **Fig. S14(B)** and its inset. It is clearly seen that $I_1$, $I_2$, and $I_T$ at $\omega_0$ are at zero level. As $\gamma_{tip}$ increases and brings the system into the weak-coupling regime, the field intensities at $\omega_\mp$ and $\omega_0$ coincide as was demonstrated in our experimental and theoretical results depicted in **Fig. S12**, **Fig. S13** and **Fig. 3C**. Note that the intracavity field intensity at $\omega_0$ increases from zero, when $\gamma_{tip} = 0$, gradually as the resonances of the eigenmodes approach each other until they coalesce at $\omega_0$.

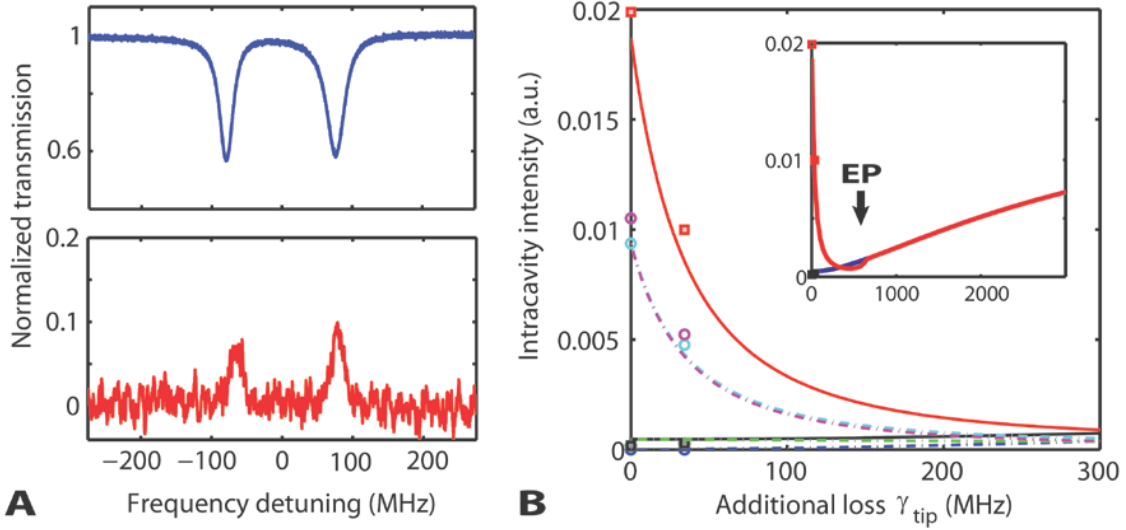

*Fig. S14. Theoretical and experimentally-obtained intracavity field intensities and experimentally-obtained transmission spectra in the very strong coupling regime. (A) Typical transmission spectra $T_{1\to 2}$ (blue) and $T_{1\to 4}$ (red) obtained in the experiments in the very strong coupling regime when $\gamma_{tip}=0$. Note that at $\omega_0$ which is between the two resonances $T_{1\to 2}\sim 1$ and $T_{1\to 4}\sim 0$ with very small noise fluctuations, implying that cavity field at $\omega_0$ is zero. (B) In the very strong coupling regime and when $\gamma_{tip}=0$, intracavity field at $\omega_0$ is zero. Theoretical intracavity field intensities $I_1, I_2$ and $I_T = I_1 + I_2$ at $\omega_\mp$ are given as dashed-dotted purple, dashed-dotted light blue, and red solid curves, respectively. Those at $\omega_0$ are given, respectively, as dashed-dotted blue, dashed-dotted green, and black solid curves. The data labeled with circles and squares were obtained from experiments. Inset shows the theoretical curves of $I_T$ at $\omega_\mp$ (red) and at $\omega_0$ (blue) showing that in this very strong-coupling regime when $\gamma_{tip}=0$ the field at $\omega_0$ is zero. Data labeled with squares are obtained from experiments. Inset of (B) shows the total intracavity field intensities at $\omega_\mp$ (red) and at $\omega_0$ (blue) for a larger range of $\gamma_{tip}$.*

## S7. Do EP and the minimum of $I_T$ always appear at different $\gamma_{tip}$?

In **Figs. 3A-3C** of the main text and **Figs. S7-S14** of this Supplement, we see that $\gamma_{tip}^{min}$, at which the total intracavity field intensity $I_T = |A_1|^2 + |A_2|^2$ takes its minimum value, differs from the $\gamma_{\text{tip}}^{\text{EP}}$ which brings the system to the EP (i.e., $\gamma_{\text{tip}}^{\min} < \gamma_{\text{tip}}^{\text{EP}}$). Using the expressions given in Eq. (S.5) and assuming $\Delta_1 = \Delta_2 = \Delta$ (i.e., $\omega_2 = \omega_1 = \omega_0$ and $\Delta = \omega - \omega_0$) as this is the condition in our experiments, we find that $I_T$ takes its minimum value at $\gamma_{\text{tip}}^{\min}$:

$$\gamma_{\text{tip}}^{\min} = \frac{\gamma_1'(\gamma_1' - 2\gamma_2') + 4(3\Delta^2 - \kappa^2) + \left[(\gamma_1'^2 + 4\kappa^2 - 12\Delta^2)^2 + 64\gamma_1'^2 \Delta^2\right]^{1/2}}{2\gamma_1'}. \quad \text{(S.12)}$$

On the other hand, from Eq. (S.3) and the discussions in the previous subsections, we know that $\gamma_{\text{tip}}$ required to bring the system to the EP is given by

$$\gamma_{\text{tip}}^{\text{EP}} = -(\gamma_2' - \gamma_1') \mp 4\kappa, \quad \text{(S.13)}$$

which, contrary to $\gamma_{\text{tip}}^{\min}$, does not depend on $\Delta$. From Eqs. (S.12) and (S.13), we find $\gamma_{\text{tip}}^{\min} = \gamma_{\text{tip}}^{\text{EP}}$ is satisfied only when Eqs. (S.14) and (S.15) are simultaneously satisfied

$$\gamma_{\text{tip}}^{\text{EP}} = -(\gamma_2' - \gamma_1') + 4\kappa; \quad \text{(S.14)}$$

$$\Delta = \frac{\sqrt{\kappa}}{2} \frac{(\gamma_1' + 2\kappa)}{\sqrt{\gamma_1' + 3\kappa}}. \quad \text{(S.15)}$$

In **Figs. S7-S14,** we presented $I_T$ at frequencies $\omega = \omega_0$ (i.e., $\Delta = 0$), $\omega = \omega_+$ (i.e., $\Delta = \omega_+ - \omega_0$), and $\omega = \omega_-$ (i.e., $\Delta = \omega_0 - \omega_-$). In none of these cases are Eqs. (S.14) and (S.15) satisfied. For example, for the first case $\omega = \omega_0$, we have $\Delta = 0$ which is satisfied when $\kappa = 0$ or $\kappa = -\gamma_1'/2$. The former implies that there is no coupling which is not the case here. The latter implies gain in the first resonator which is not realized in our experiments. (Note that $\gamma_1' > 0$ implies net loss whereas $\gamma_1' < 0$ implies net gain in the first resonator). In addition, plugging $\kappa = -\gamma_1'/2$ in Eq. (S.14) leads to $\gamma_{\text{tip}}^{\text{EP}} = -\gamma_2' + \gamma_1' - 2\gamma_1' = -(\gamma_2' + \gamma_1')$ which implies that under this condition the EP can only be observed with a net total gain in the compound resonator

S25

system. Therefore, it is normal that $\gamma_{tip}^{min}$ and $\gamma_{tip}^{EP}$ do not coincide in our simulations and experiments where we have calculated and estimated $I_T$ at $\omega = \omega_0$. In **Fig. S15**, we present the effect of $\Delta$ on $\gamma_{tip}^{min}$ both theoretically and experimentally, and determine $\Delta$ which leads to $\gamma_{tip}^{min} = \gamma_{tip}^{EP}$ in our system. The theoretical curves were plotted for typical values of system parameters in our experiments. As $\Delta$ increases, $\gamma_{tip}^{min}$ approaches $\gamma_{tip}^{EP}$. The theoretically expected value for $\Delta$ leading to $\gamma_{tip}^{min} = \gamma_{tip}^{EP}$ in our system is $\Delta = 19.8$ MHz which is in good agreement with what we observe in the experiments. The slight discrepancy between theory and experiment is within our experimental error range and can be attributed to the shallowness of the minimum as well as to accumulated errors in curve fittings, loss of phase information in the measurement of transmission spectra between ports 1 and 2, and to the frequency fluctuations of the probe laser. Note that we do not use any active or passive stabilization or locking methods in our experiments.

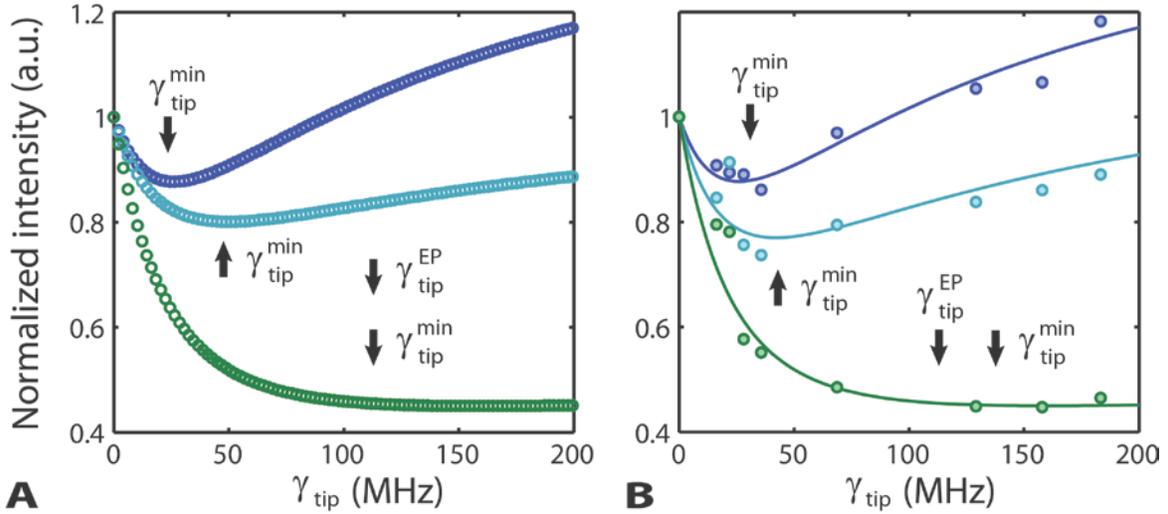

*Fig. S15. Effect of the frequency detuning $\Delta$ from the exceptional point (EP) frequency $\omega_0$ on the additional loss $\gamma_{tip}^{min}$ at which total intracavity field intensity $I_T$ reaches its minimum. Normalization was done with the total intensity at $\gamma_{tip} = 0$. (A) Theoretical and (B) experimentally-obtained total intracavity field intensity $I_T$ as a function of $\gamma_{tip}$. In (B) circles are experimentally-obtained data whereas the lines are from the theoretical model. Blue, cyan and green colors correspond, respectively, to $\Delta = 0$, $\Delta = 8$ MHz, and $\Delta = 19.8$ MHz.*

## S8. Thermal response of coupled-resonator system.

In **Fig. 3D** and the related parts of the main text, we discussed thermal nonlinearity as a manifestation of enhanced field intensity build-up within WGM resonators. In the experiments reported in **Fig. 3D**, we used the configuration shown in **Fig. S1(C)** where only μR$_1$ is coupled to a fiber-taper coupled WG1 (that is, WG2 is far away from μR$_2$ so that there is no coupling between them). This configuration can be modeled by setting $\gamma_{c2} = 0$ in Eqs. (S.1) and (S.2). Thus, the thermal response of the coupled resonator system is characterized by the following set of equation *(28,33,34)*:

$$\frac{da_1}{dt} = -i\Delta_1 a_1 - \frac{\gamma_1 + \gamma_{c1}}{2} a_1 - i\kappa a_2 - \sqrt{\gamma_{c1}} a_{in}$$
(S.16)

$$\frac{da_2}{dt} = -i\Delta_2 a_2 - \frac{\gamma_2 + \gamma_{tip}}{2} a_2 - i\kappa a_1$$

$$\frac{dT_1(t)}{dt} = -\alpha_{abs} \frac{|a_1|^2}{\tau_{r1}} + \alpha_{th} T_1(t)$$
(S.17)

$$\frac{dT_2(t)}{dt} = -\alpha_{abs} \frac{|a_2|^2}{\tau_{r2}} + \alpha_{th} T_2(t)$$

where Eq. (S.16) describes the time evolution of the intracavity optical fields of the coupled resonators, and Eq. (S.17) describes the evolution of the temperature inside the mode volumes of the microresonators. $\alpha_{abs}$ is the thermal absorption coefficient and is responsible for the temperature change due to the absorption of light by the material used to fabricate the microresonator (in our case, silica). $\alpha_{th}$ represents the thermal relaxation rate and quantifies the heat dissipation process. $T_{j=1,2}(t)$ is the temperature inside the resonator.

In our experiments, the wavelength of a tunable laser is scanned to probe the resonances and response of the microresonators. As the laser wavelength is up-scanned from shorter to longer wavelengths to approach the resonance wavelength (decreasing detuning), the intracavity fields are gradually built up inside the resonators. Material absorption then gives rise to an increase in the temperature of the resonator. This, on the other hand, alters the refractive index of the resonators through the thermo-optic effect and therefore shifts the resonance frequencies of the resonators. In order to take this dynamic effect into account and complete the model, we introduce the following two equations:

$$\begin{aligned}\Delta\omega_1 &= \omega(t) - \omega_1\left[1 - a\times dT_1(t)\right]\\ \Delta\omega_2 &= \omega(t) - \omega_2\left[1 - a\times dT_2(t)\right]\end{aligned} \quad (S.18)$$

where $a$ represents a thermal coefficient which takes into account the thermal expansion and the thermo-optic coefficients.

*Table S1. Values of the parameters used in the numerical simulations for the thermal response of the coupled resonators.*

| Parameter | Value (Case 1) | Value (Case 2) | Unit |
|---|---|---|---|
| **Intrinsic loss of µR1: $\gamma_1$** | 24.18 | 9.67 | MHz |
| **Intrinsic loss of µR2: $\gamma_2$** | 19.34 | 38.68 | MHz |
| **Additional loss: $\gamma_{tip}$** | 77.38<br>947.75 | 90.26<br>928.41 | MHz |
| **Coupling loss between WG1 and µR1: $\gamma_{c1}$** | 24.18 | 24.18 | MHz |
| **Inter-resonator coupling strength: $\kappa$** | 36.27 | 14.51 | MHz |
| **Resonant wavelength of solitary resonators: $\lambda_0$** | 1550 | 1550 | nm |
| **Thermal coefficient: $a$** | $6\times10^{-6}$ | $6\times10^{-6}$ | 1 / °C |
| **Thermal relaxation rate: $\alpha_{th}$** | 90 | 90 | kHz |
| **Thermal absorption coefficient: $\alpha_{abs}$** | $1.83\times10^3$ | $1.83\times10^3$ | K/J |

We numerically solved Eqs. (S.16)-(S.18) to characterize the thermal response of the coupled resonators used in our experiments, and to quantify how well this theoretical model represents our experimental results shown in **Fig. 3D** of the main text. We performed numerical simulations for the Case 1 and Case 2 with the same setting [**Fig. S1(C)**] used in **Fig. 3D**. These two cases differ in whether the additional loss $\gamma_{tip}$ is introduced to the resonator with initially lower or higher quality factor. In **Table S1**, we have listed the value of the parameters used in the simulations. The results are shown in **Fig. S16(A)** and **Fig. S16(C)** which clearly show that with increasing additional loss the thermal response of the coupled-resonators system evolves into a waveform similar to that of a single resonator [top spectra in **Fig. S16(A)** and **Fig. S16(C)**]. The reason for this can be seen in the evolution of the intracavity field intensities [**Fig. S16(B)** and Fig. **S16(D)**]. As the additional loss is increased the field becomes localized in only one resonator, thus thermally-affecting only this resonator (the effect of the other resonator on the thermal response spectra is negligible). The recovery of the thermal response is not perfect because field localization in only one resonator is not perfect (less than 100%), that is, there still exists field in the other resonator. Mode splitting seen in **Fig. S16(A)** is due to the strong coupling of the resonators which overcomes the total loss in the system. This manifests itself in **Fig. S16(B)** as almost equal intracavity field intensities (i.e., in the strong coupling regime the field is distributed in both of the resonators equally). As the loss is increased, the system moves into the weak-coupling regime and the splitting is lost. Consequently, the field is localized in the resonator with less loss. In **Fig. S16(C)**, we do not see mode-splitting because the system is already close to the weak-coupling regime, and a small splitting is buried within the thermally broadened transmission spectra. In this regime, as expected, the intracavity field intensities of the resonators are different, and the difference increases with increasing additional loss.

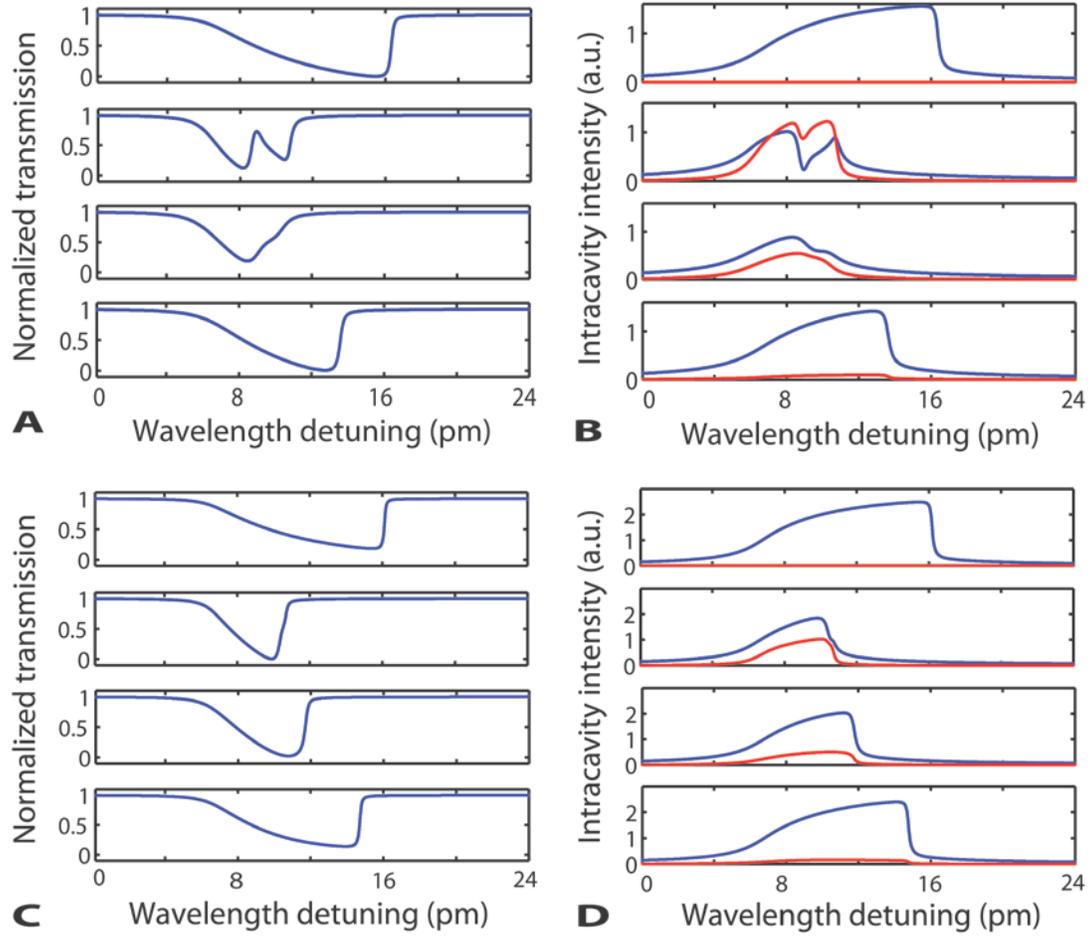

*Fig. S16. Theoretically-obtained thermal response of coupled resonators.* *(A)* and *(B)* Normalized transmission spectra and the corresponding intracavity field intensity spectra as the loss induced in the second microresonator is increased from top to bottom. The initial loss of the first resonator is higher than that of the second resonator where the additional loss $\gamma_{tip}$ is introduced (Case 1). *(C)* and *(D)* Normalized transmission spectra and the corresponding intracavity field intensity spectra as the loss induced in the second microresonator is increased from top to bottom. Initial loss of the first resonator is lower than that of the second resonator where $\gamma_{tip}$ is introduced (Case 2). The values of parameters used in the simulations are listed in Table S1. In (B) and (D), blue and red curves denote the intracavity field intensities in the first and second resonator, respectively. In (A) and (C), only the transmission spectrum for $T_{1\rightarrow 2}$ is depicted. Top panel in (A)-(D) is obtained when the second resonator is far away from the first one such that there is no coupling between the resonators.

**References:**

1. C. M. Bender, Making sense of non-Hermitian Hamiltonians, *Rep. Prog. Phys.* **70**, 947 (2007).

2. I. Rotter, A non-Hermitian Hamilton operator and the physics of open quantum systems, *J. Phys. Math. Theor.* **42**, 153001 (2009).

3. N. Moiseyev, Non-Hermitian Quantum Mechanics, Cambridge University Press (2011).

4. W. D. Heiss, Exceptional points of non-Hermitian operators, *J. Phys. A* **37**, 2455 (2004).

5. E. Persson, I. Rotter, H. J. Stöckmann, M. Barth, Observation of resonance trapping in an open microwave cavity, *Phys. Rev. Lett.* **85**, 2478–2481 (2000).

6. C. Dembowski et al., Experimental observation of the topological structure of exceptional points, *Phys. Rev. Lett.* **86**, 787–790 (2001).

7. S. B. Lee et al., Observation of an exceptional point in a chaotic optical microcavity, *Phys. Rev. Lett.* **103**, 134101 (2009).

8. C. M. Bender, S. Boettcher, Real spectra in non-Hermitian Hamiltonians having PT symmetry, *Phys. Rev. Lett.* **80**, 5243–5246 (1998).

9. R. El-Ganainy, K. G. Makris, D. N. Christodoulides, Z. H. Musslimani, Theory of coupled optical PT-symmetric structures, *Opt. Lett.* **32**, 2632 (2007).

10. C. E. Rüter et al., Observation of parity–time symmetry in optics, *Nat. Phys.* **6**, 192–195 (2010).

11. A. Guo et al., Observation of PT-symmetry breaking in complex optical potentials, *Phys. Rev. Lett.* **103**, 093902 (2009).

12. A. Regensburger et al., Parity-time synthetic photonic lattices, Nature 488, 167–171 (2012).

13. L. Feng et al., Experimental demonstration of a unidirectional reflectionless parity-time metamaterial at optical frequencies, *Nat. Mater.* **12**, 108–113 (2013).

14. L. Feng et al., Demonstration of a large-scale optical exceptional point structure, *Opt. Express* **22**, 1760–1767 (2014).

15. H. Wenzel, U. Bandelow, H.-J. Wunsche, J. Rehberg, Mechanisms of fast self pulsations in two-section DFB lasers, *IEEE J. Quantum Electron.* **32**, 69 –78 (1996).

16. M. V. Berry, Mode degeneracies and the Petermann excess-noise factor for unstable lasers, *J. Mod. Opt.* **50**, 63–81 (2003).

17. M. Liertzer et al., Pump-induced exceptional points in lasers, *Phys. Rev. Lett.* **108**, 173901 (2012).

18. B. Peng et al., Parity-time-symmetric whispering-gallery microcavities, *Nat. Phys.* **10**, 394–398 (2014).

19. M. Brandstetter et al., Reversing the pump-dependence of a laser at an exceptional point, *Nat. Commun.* **5**, 4034 (2014).

20. Supplementary materials are available on Science Online.

21. K. J. Vahala, Optical microcavities. *Nature* **106**, *839-846 (2003)*.

22. D. O'Shea, C. Junge, J. Volz, and A. Rauschenbeutel, Fiber-optical switch controlled by a single atom, *Phys. Rev. Lett.* **111**, 193601 (2013).

23. T. J. Kippenberg, K. J. Vahala, Cavity optomechanics: Back-action at the mesoscale, *Science* **321**, 1172-1176 (2008).

24. L. He, S. K. Ozdemir, L. Yang, Whispering gallery microcavity lasers, *Laser &Photon. Rev.* **7**, 60 (2013).

25. X. Fan, I. M. White, S. I. Shopova, H. Zhu, J. D. Suter, Y. Sun, Sensitive optical biosensors for unlabeled targets: A review, *Anal. Chim. Acta*. **620**, 8– 26 (2008).

26. F. Vollmer, S. Arnold, Whispering-gallery-mode biosensing: label-free detection down to single molecules, *Nat. Meth.* **5**, 591–596 (2008).

27. J. Zhu *et al.,* Single nanoparticle detection and sizing by mode-splitting in an ultra-high-*Q* microtoroid resonator, *Nat. Photon.* **4**, 46-49 (2010).

28. T. Carmon, L. Yang, K. J. Vahala, Dynamical thermal behavior and thermal self-stability of microcavities, *Opt. Express* **12**, 4742–4750 (2004).

29. S. Spillane, T. J. Kippenberg, K. J. Vahala, Ultralow-threshold Raman laser using a spherical dielectric microcavity, *Nature* **415**, 621 –623 (2002).

30. S. K. Ozdemir *et al.,* Highly sensitive detection of nanoparticles with a self-referenced and self-heterodyned whispering-gallery Raman microlaser, *Proc. Natl. Acad. Sci. USA*, **111**, E3836-E3844 (2014).

31. C. W. Gardiner and P. Zoller, Quantum Noise: A Handbook of Markovian and Non-Markovian Quantum Stochastic Methods with Applications to Quantum Optics, Springer-Verlag Berlin Heidelberg (2010).

32. D. K. Armani, T. J. Kippenberg, S. M. Spillane, K. J. Vahala, Ultra-high-Q toroid microcavity on a chip, *Nature* **421**, 925-928 (2003).

33. L. He, Y. Xiao, J. Zhu, S. Ozdemir, and L. Yang, Oscillatory thermal dynamics in high-Q PDMS-coated silica toroidal microresonators, *Opt. Express* **17**, 9571-9581 (2009).

34. I. Grudinin and K. Vahala, Thermal instability of a compound resonator, *Opt. Express* **17**, 14088-14097 (2009).